\documentclass[a4paper,fleqn,usenatbib]{mnras}

\usepackage{newtxtext,newtxmath}

\usepackage{multirow}
\usepackage{makecell}

\usepackage[T1]{fontenc}
\usepackage{array}
\DeclareRobustCommand{\VAN}[3]{#2}
\let\VANthebibliography\thebibliography
\def\thebibliography{\DeclareRobustCommand{\VAN}[3]{##3}\VANthebibliography}

\usepackage{graphicx}	
\usepackage{amsmath}	

\usepackage{url}
\usepackage{hyperref}

\usepackage[usenames,dvipsnames]{xcolor}

\title[Cosmic growth rate from the homogeneity scale]
{The homogeneity scale and the growth rate of  cosmic structures}

\author[F. Avila et al.]{
Felipe Avila,$^{1}$\thanks{E-mail: felipeavila@on.br}
Armando Bernui,$^{1}$
Rafael C. Nunes,$^{2}$
Edilson de Carvalho,$^{3}$ 
Camila P. Novaes$^{2}$\\
$^{1}$Observat\'orio Nacional, Rua General Jos\'e Cristino 77, 
S\~ao Crist\'ov\~ao, 20921-400 Rio de Janeiro, RJ, Brazil \\
$^{2}$Instituto Nacional de Pesquisas Espaciais, Av. dos Astronautas 1758, 
Jardim da Granja, S\~ao Jos\'e dos Campos, SP, Brazil \\
$^{3}$Centro de Estudos Superiores de Tabatinga, Universidade do Estado do Amazonas, 
69640-000, Tabatinga, AM, Brazil
}

\date{Accepted XXX. Received YYY; in original form ZZZ}

\pubyear{2021}

\begin{document}
\label{firstpage}
\pagerange{\pageref{firstpage}--\pageref{lastpage}}
\maketitle

\begin{abstract}
We propose a novel approach to obtain the growth rate of cosmic structures, $f(z)$, 
from the evolution of the cosmic homogeneity scale,  $R_{\text{H}}(z)$. 
Our methodology needs two ingredients in a specific functional form: 
$R_{\text{H}}(z)$ data and the matter two-point correlation function today, i.e., $\xi(r, z=0)$.
We use a Gaussian Process approach to reconstruct the function $R_{\text{H}}$. 
In the absence of suitable observational information of the matter correlation function in 
the local Universe, $z \simeq 0$, we assume a fiducial cosmology to obtain  $\xi(r, z=0)$.
For this reason, our final result turns out to be a consistency test of the cosmological 
model assumed. 
Our results show a good agreement  between: (i) the growth rate $f^{R_{\text{H}}}(z)$ 
obtained through our approach, (ii) the $f^{\Lambda\text{CDM}}(z)$ expected in the fiducial model, and (iii) the best-fit $f(z)$ from data compiled in the literature. 
Moreover, using this data compilation, we perform a Gaussian Process to reconstruct 
the growth rate function $f^{\text{data}}(z)$ and compare it with the function $f^{R_{\text{H}}}(z)$ finding a concordance of $< \!2 \,\sigma$, a good result considering the few data 
available for both reconstruction processes. 
With more accurate $R_{\text{H}}(z)$ data, from forthcoming surveys, the homogeneity scale 
function might be better determined and would have the potential to discriminate between 
$\Lambda$CDM and alternative scenarios as a new cosmological observable.
\end{abstract}



\section{Introduction}\label{introduction}

There is an increasing interest in measurements of the growth rate of cosmic structures, 
$f(z)$, because this function behaves differently for cosmological models based on different 
theories of gravity (see, e.g.,~\cite{Huterer15, Kasa18, Basilakos20, Linder20, Velasquez20}); 
notoriously, the concordance cosmological model $\Lambda$CDM  is based on the 
theory of general relativity. 
In this scenario, precise measurements of $f(z)$ from diverse cosmological tracers 
measured at several redshifts would determine if the $\Lambda$CDM model correctly 
describes the evolution of the function $f(z)$~\citep{Pezotta,Aubert20,Bautista21,Avila21}, 
and to investigate classes of models based on modified gravity theory~\citep{Alam20,Ntelis20}.
But the interest in $f(z)$ is more fundamental. 
In fact, since the early works of~\citet{Peebles65,Silk68,Sunyaev}, the theory of cosmological 
perturbations searches to describe the clustering evolution of the primordial density 
fluctuations, from the earliest times to the currently observed universe, where the growth rate 
of structures $f(z)$ represents a measurement of such clustering evolution. 

The measurements of $f(z)$ can be done with good precision using the Redshift Space 
Distortions (RSD) approach, that is, studying the peculiar velocities caused by local 
gravitational potentials that introduce distortions in the two-point correlation function 
(2PCF)~\citep{Kaiser87}. 
Calculating the 2PCF from a galaxy survey, more precisely, the anisotropic correlation function, 
$\xi(s, \mu)$~\citep{Hamilton92,Hamilton95}, one can constrain the product $f \sigma_{8}$, 
where $\sigma_{8}$ is the variance of the matter fluctuations at the scale of 
8~Mpc$/h$~\citep{Juszkiewicz,Song09}. For $f\sigma_{8}$ data compilations see, e.g.,~\cite{Zhang,Sagredo18,Alam21}.

The growth rate of cosmic structures, $f$, is defined as \citep{Strauss}
\begin{equation}\label{GR}
f(a) \equiv \dfrac{d\ln D(a)}{d\ln a} \, , \nonumber
\end{equation}
where $D = D(a)$ is the linear growth function, and $a$ is the scale factor in the 
Robertson-Walker metric, based on general relativity theory. 
Apply the above equation to a catalogue of cosmic objects to measure $f$ does not work, because  what one can measure directly from the data survey is the density 
contrast $\delta(r, a)$ and not the growth function $D(a)$. 
In this work we propose a solution for this problem: 
search for a cosmic observable function that depends only on cosmic time 
(equivalently, on the redshift $z$ or the scale factor $a$) instead of 
$D(a)$ in the above 
equation, being able to quantify the clustering evolution to  provide a measurement of 
the growth rate of cosmic structures.

The Cosmological Principle is a fundamental piece in the concordance model of 
cosmology~\citep{peebles80}. 
It claims that, at sufficiently large scales, the universe is statistically homogeneous and 
isotropic 
\citep[regarding the statistical isotropy of the universe 
see, e.g.,][]{Bernui07,Bernui14,Pereira15,Tarnopolski17,Bengaly17,Dainotti18,Marques18,Shafieloo}. 
Several teams analysed galaxy surveys to calculate the scale where the transition from 
an inhomogeneous to a homogeneous distribution occurs, termed the homogeneity scale 
$R_{\text{H}}$~\citep{Scrimgeour,Laurent16,Ntelis17}. For recent analyses 
see, e.g., \cite{Ntelis19, Heinesen20,Pandey21a,Pandey21b,Rodrigo21,deMarzo21,Camacho21}.
In addition, analyses of the angular scale homogeneity have also been done to find the angular 
scale of homogenenity, $\theta_{\text{H}}$,~\citep{Alonso15,Rodrigo18a,Avila18,Avila19}, 
considered model independent measurements because one does not assume a cosmological model, 
as in the analyses of $R_{\text{H}}$, when one uses a fiducial cosmology to calculate 3D distances. 
At present, diverse deep astronomical surveys map large volumes of the universe, permitting 
to probe the evolution of $R_{\text{H}}$, although it is not as accurate as desirable. 
The next generation of surveys foresees a large number of $R_{\text{H}}$ 
measurements with an improved accuracy~\citep{Amendola18,Ivezic19}.

In this work we will show that it is possible to use information from $R_{\text{H}}$, more 
precisely from the homogeneity scale evolution $dR_{\text{H}}/dz$, to obtain the cosmic 
evolution of the growth rate of structures $f(z)$. 
From the theoretical point of view, the homogeneity scale can be related to the 2PCF, 
$\xi(r)$~\citep{peebles80,Ntelis17}. 
From the linear perturbation theory, the redshift evolution of $\xi(r)$ is proportional to 
$D(z)^{2}$ then, $F[R_{\text{H}}(z)] D(z)^{2}\propto \text{cte}$, where $F$ is a functional 
of the homogeneity scale function $R_{\text{H}}(z)$. 
As we shall see, this proportionality leads to the growth rate $f(z)$ through the redshift 
derivative of $\bar{\xi}(R_{\text{H}}(z))$, the volume-averaged 2PCF. 
The approach to know the functional $F$ needs to assume parameters that we determine 
assuming a $\Lambda$CDM fiducial cosmology. 
In this sense, our analyses should be considered as a test of consistency for the 
$\Lambda$CDM model.

The relationship between $f$ and $R_{\text{H}}$ indicates that with precise homogeneity 
scale data, $R_{\text{H}}(z)$, measured at several redshifts, one can determine 
with good accuracy the growth rate of cosmic structures $f=f(z)$, which in turn can be 
used to discriminate between the concordance $\Lambda$CDM and competing models based 
on modified gravity theories. 
Additionally, these data could be used in statistical analyses to find cosmological parameters. 
In other words, the homogeneity scale data, $R_{\text{H}}(z)$, would indeed play the role of a 
novel cosmological observable, as first discussed by~\citet{Ntelis19}.

This work is organized as follows. In section \ref{growthratesection} we review the main equations of the linear theory of matter perturbations. In section \ref{transitionscaletohomogeneity} we explain the methodology to obtain the transition scale to homogeneity and, for the first time, the relation between $R_{\text{H}}(z)$ and $f(z)$. In section \ref{testingthemodel} we explain the reconstruction procedure to obtain a smooth curve of $R_{\text{H}}(z)$, and  $dR_{\text{H}}(z)/dz$, 
both used then to obtain  $f(z)$ according to our procedure. 
In sections \ref{resulsanddiscussions} we show our results and discuss them,  while in section \ref{conclusions} we present our conclusions.

\section{Growth Rate of Cosmic Structures}\label{growthratesection}

To describe the structure formation in an isotropic and homogeneous universe we used a 
perturbation approach: small deviation in the early universe has a slow evolution that can 
be described by a linear perturbation theory~\citep{Mukhanov}. 
One defines the density contrast as 
\begin{equation}\label{constrastdensity}
\delta(\textbf{r}, t) \equiv \dfrac{\rho(\textbf{r}, t)-\bar{\rho}(t)}{\bar{\rho}(t)} \, ,
\end{equation}
where $\rho(\textbf{r}, t)$ is the matter density at the comoving vector position \textbf{r} at cosmic time 
$t$ and $\bar{\rho}(t)$ is the average matter density measured in the 
hyper-surface of constant $t$. 
In the linear and Newtonian regime, the gravitational potentials are small and 
the perturbation scale is smaller than the Hubble radius, $\lambda \ll c/H_{0}$, where 
$c$ is the speed of light and $H_{0}$ is the Hubble constant. 
Over this condition, the structure formation is described with the fluid equations
\begin{equation}\label{continuity}
\dot{\delta} = -\frac{1}{a}\nabla\cdot\textbf{v} \, ,
\end{equation}
\begin{equation}\label{Euler}
\dot{\textbf{v}} + H\textbf{v} = -\frac{1}{a\bar{\rho}}\nabla\delta p - \frac{1}{a}\nabla\delta\Phi \, ,
\end{equation}
\begin{equation}\label{Poisson}
\nabla^{2}\delta\Phi = 4\pi Ga^{2}\bar{\rho}\,\delta \, ,
\end{equation}
which are the continuity, Euler, and Poisson equations, respectively, perturbed at first order in comoving space. The dot corresponds to a partial derivative in cosmic time. 
The physical quantities \textbf{v}, $\delta p$, and $\delta\Phi$ are the peculiar velocity, pressure, and the perturbed gravitational potential, respectively.

Combining equations (\ref{continuity}), (\ref{Euler}), and (\ref{Poisson}), and assuming adiabatic perturbations condition, we obtain the well known equation that describes the linear density contrast evolution of the matter density

\begin{equation}\label{mat.dominante}
\ddot{\delta}_{m} + 2\frac{\dot{a}}{a}\delta_{m} - 4\pi G\bar{\rho}_{m}\delta_{m} = 0 \, .
\end{equation}
In the linear approximation, the density contrast is a function of time only, that is, 
$\delta_{m}\sim D(t)$. 
From this, one can define the growth rate of cosmic structures as
\begin{equation}\label{growthrate}
f(a)\equiv \frac{a}{D} \frac{dD}{da} = \dfrac{d\ln D}{d\ln a} \, .
\end{equation}
In the $\Lambda$CDM model we have the following  approximation~\citep{Lahav91}
\begin{equation}\label{fapprox}
f(z) \,\simeq\, \Omega_{m}^{0.6}(z) \,+\, \frac{\Omega_{\Lambda}}{70} 
\left(1 + \frac{1}{2} \Omega_{m}(z) \right) \, ,
\end{equation}
where $\Omega_{m}$ and $\Omega_{\Lambda}$ are the matter and dark energy 
cosmological parameters, respectively. 
An alternative approximation is given by~\citep{Linder05,Linder07}
\begin{equation}\label{Lindeapprox}
f(z) = \Omega_{m}^{\gamma}(z) \, ,
\end{equation}
where $\gamma$ is the growth index. In the $\Lambda$CDM model, 
$\gamma = 6/11 \simeq 0.55$. 
This parameter assumes distinct values beyond $\Lambda$CDM cosmology~\citep{Basilakos12}.

\section{Transition Scale to Homogeneity}\label{transitionscaletohomogeneity}

The most used methodology to study the homogeneity of galaxy or quasars  distributions is to count the number of cosmic  objects, $N_{\text{gal}}$, inside a sphere of radius $r$, and divide for $N_{\text{rand}}$, the equivalent count but for a random distribution, that has the same features as the original one. 
Then, we can define the {\em scaled counts-in-spheres}, 
$\mathcal{N}(< r)$~\citep{Scrimgeour} 
\begin{equation}\label{scaledcounts}
\mathcal{N}(< r) \equiv \frac{N_{\text{gal}}(< r)}{N_{\text{rand}}(< r)} \, ,
\end{equation}		 
where for a homogeneous distribution, at large scales, it goes to 1. 
It can be shown that $ \mathcal{N}(<r) $ is related to the two-point correlation function $\xi(r)$ 
\footnote{for applications of the two-point correlation function in clustering analyses see, e.g., \cite{deCarvalho18,deCarvalho21,Carvalho}.}
\begin{equation}\label{scaledcountscorrelationfunction}
\mathcal{N}(< r) = 1 + \frac{3}{r^{3}}\int_{0}^{r}\xi(s)s^{2}ds \, .
\end{equation}
From the function $\mathcal{N}(< r)$ one can define the correlation dimension function 
$\mathcal{D}_{2}(r)$ 
\citep[for details see the Appendix A in][]{Ntelis17}, 
\begin{equation}\label{cordimension}
\mathcal{D}_{2}(r) \equiv \frac{d~\ln\mathcal{N}(< r)}{d~\ln r} + 3 \, .
\end{equation}
Despite the fact that most studies present both estimators, $ \mathcal{N}(<r) $ and $ \mathcal{D}_{2}(r) $, the result from the correlation dimension is considered more robust, because it is less correlated for most scales~\citep{Scrimgeour,Ntelis17}.

To finish this section we discuss the arbitrary criterion used to determine the scale 
where the transition to homogeneity occurs, $R_{\text{H}}$. Consider the following equation
\begin{equation}\label{homogeneityscale}
\mathcal{D}_{2}(R_{\text{H}}) = 3(1 - \epsilon).
\end{equation}
In an ideal situation, one expects  $\epsilon = 0$, that is, when the homogeneity scale is attained the value of $\epsilon$ should be $\epsilon = 0$, 
such that the transition to homogeneity occurs on the scale at which $\mathcal{D}_{2}$ calculated from data achieve the value 3. 
However, due to systematic effects present in the galaxy surveys, \cite{Scrimgeour} suggested 
to fix the value of $\epsilon$ 
at, for example, 0.01, 
which gives us $\mathcal{D}_{2}(R_{\text{H}}) = 2.97$, 
that is, 1\% below 3. 
We assume this value because it is commonly adopted in the literature, and allow us to study the scale of homogeneity for different tracers in a large 
range of redshift. 
Anyhow, as we will show next, our methodology is independent of $\epsilon$, due to the 
redshift derivative.

\subsection{The growth rate of cosmic structures from the  homogeneity scale}\label{3.1}
As mentioned above, the scaled counts-in-spheres, $\mathcal{N}(<r)$, 
is related to the two-point correlation function, $\xi(r;z)$, at redshift $z$
\begin{equation}\label{correlationfunction}
\xi(r=|\textbf{x}-\textbf{y}|;z) = \langle\delta(\textbf{x};z)\delta(\textbf{y};z)\rangle \,,     
\end{equation}
that is, is the spatial average of the product of the density contrasts evaluated at the arbitrary positions of a pair of galaxies, $\textbf{x},\,\textbf{y}$, at  redshift $z$. 
The redshift evolution of $\xi$ can be obtained assuming for the equation (\ref{mat.dominante}) the solution $\delta(r;z) = \delta(r; z=0)D(z)$~\citep{Schneider}. This leads to
\begin{equation}\label{correlationfunctionz}
\begin{split}
\xi(r=|\textbf{x}-\textbf{y}|; z) &= \langle\delta(\textbf{x}; z) \delta(\textbf{y}; z)\rangle \\ 
 &= D^{2}(z)\langle\delta(\textbf{x};z=0)\delta(\textbf{y};z=0)\rangle \\
 &= D^{2}(z)\xi(r;z=0) \, ,
\end{split}
\end{equation}
where $\xi(r; z=0)$ is the two-point correlation function at $z=0$. From equation (\ref{correlationfunctionz}) one  can rewrite the scaled counts-in-spheres as
\begin{equation}\label{N_r_z}
\mathcal{N}(< r, z) = 1 + D^{2}(z)\bar{\xi}(r) \, ,
\end{equation}
where
\begin{equation}\label{xibarra}
\bar{\xi}(r) \equiv  \frac{3}{r^{3}}\int_{0}^{r}\xi(s,z=0)s^{2} ds \, ,
\end{equation}
is the volume average of the correlation function. 
From the equation (\ref{cordimension}) one has 
\begin{equation}\label{D2_r_z}
\mathcal{D}_{2}(r, z) 
= \frac{rD^{2}(z)}{1 + D^{2}(z)\bar{\xi}(r)}\frac{d\bar{\xi}(r)}{dr} + 3 \, .
\end{equation}
It is useful to define the following quantity
\begin{equation}\label{chiequation}
\zeta(r) \equiv \frac{d\bar{\xi}(r)}{dr} \, .
\end{equation}
For the scale where the transition to homogeneity occurs, $r = R_{\text{H}}$, 
equation (\ref{D2_r_z}) becomes 
\begin{equation}\label{Rhequation}
R_{\text{H}}(z)D^{2}(z)\zeta[R_{\text{H}}(z)] 
= - 3 \epsilon\,(1 + D^{2}(z)\,\bar{\xi}[R_{\text{H}}(z)]) \simeq - 3 \epsilon\, ,
\end{equation}
where we consider only the first-order  term. 
Now, taking the redshift derivative of equation (\ref{Rhequation}) we have
\begin{equation}\label{Rhderivation}
\frac{d}{dz}(R_{\text{H}}D^{2}\zeta) = 0 \, .
\end{equation}
This differential equation relates in a simple way $R_{\text{H}}(z)$ and $D(z)$.

Finally, separating each term of equation (\ref{Rhderivation}) we have
\begin{equation}\label{eq1}
-\frac{2}{D}\frac{dD}{dz} = \frac{1}{R_{\text{H}}}\frac{dR_{\text{H}}}{dz} + \frac{1}{\zeta}\frac{d\zeta}{dz}. 
\end{equation}
Using the equation (\ref{growthrate}) in equation (\ref{eq1}), we have 
\begin{equation}\label{fz_Rh}
f(z) = \frac{1+z}{2}\left(\frac{1}{R_{\text{H}}}\frac{dR_{\text{H}}}{dz} + \frac{1}{\zeta}\frac{d\zeta}{dz}\right),
\end{equation}
which, explicitly, is independent of $\epsilon$. 
To obtain $f(z)$, in addition to $R_{\text{H}}$ data, we must obtain 
$\zeta[R_{\text{H}}(z)]$ from a correlation function in $z=0$. 
Or, in a model dependent way, use an approximation, as we will describe in the next 
section.

\section{Testing the Model}\label{testingthemodel}

In this section, 
we describe our methodology, aimed to solve equation (\ref{fz_Rh}), and apply it to 
a set of $R_{\text{H}}$ data to 
obtain the growth function $f(z)$. 
Firstly, we present the data, and secondly, we detail the approximation used to define 
$\zeta(R_{\text{H}})$. 
By last, we describe the Gaussian Process methodology used to reconstruct $R_{\text{H}}$ 
and $dR_{\text{H}}/dz$.

\subsection{Data}\label{data}


Here, we use two sets of $R_{\text{H}}$ measurements. 
The first one is provided by~\cite{Ntelis17}, through the study of the CMASS galaxy 
sample of the BOSS survey, they calculated the transition to homogeneity for 5 uncorrelated redshift bins in the interval $0.43-0.70$. 
The authors analysed 
separately
the North (NGC) and South Galactic Caps (SGC), at the same redshift bins, obtaining 5 independent measurements for each of them (i.e., a total of 10 $R_{\text{H}}$ data). 
The second $R_{\text{H}}$ data set comes from \cite{Rodrigo18b}, who analysed the quasars sample from the fourteenth data release of the Sloan Digital Sky Survey (SDSS-IV DR14) in the redshift interval $0.80 - 2.24$.
They measured $R_{\text{H}}$ in each one of 4 uncorrelated redshift bins (i.e., 4 $R_{\text{H}}$ data) studied employing two estimators to calculate  $\mathcal{N}(<r)$: 
Landy-Szalay (LS) and Peebles-Hauser (PH) estimators, obtaining similar results in both  cases. 
In Table~\ref{tab:Rhdatapoints} we list these $R_{\text{H}}(z)$ data with their respective redshifts. 

Following \cite{Ntelis19}, we combine the $R_{\text{H}}$ data from NGC ($j$) and SGC ($k$) using a weighted average, defined as
\begin{equation}\label{weighted_data}
R_{\text{H}}^{\text{w}}(z_{i}) \equiv \left(\frac{1}{\sigma_{j}^{2}(z_{i})} + \frac{1}{\sigma_{k}^{2}(z_{i})}\right)^{-1}\times \left(\frac{R_{\text{H}}^{j}(z_{i})}{\sigma_{j}^{2}(z_{i})} + \frac{R_{\text{H}}^{k}(z_{i})}{\sigma_{k}^{2}(z_{i})}\right),
\end{equation}
where $\sigma_{j}$, $\sigma_{k}$, and $R_{\text{H}}^{j}$, $R_{\text{H}}^{k}$ are, respectively, the errors and data for two independent measurements, $j$ and $k$, in the same redshift, $z_{i}$. 
At first order, we can neglect the covariance between redshift bins. For the~\cite{Rodrigo18b} data, we choose the data from the PH estimator to optimize our analyses.


Notice that the $R_{\text{H}}$ we use here are already corrected by the corresponding bias factor. 
The bias for each redshift bin have been determined by the respective authors, which provided the bias-corrected measurements.
This is important because each homogeneity scale measurement is calculated for a specific tracer and a proper combination of these data requires their conversion to the corresponding transition scale for the underlying matter distribution.

\begin{table}
	\linespread{1.2}
	\selectfont
	\centering
	\caption{The $R_{\text{H}}(z)$ data used in the analyses.}
	\label{tab:Rhdatapoints}
	\begin{tabular}{c c c c l}
		\hline
		$z$ & \multicolumn{2}{c}{$R_{\text{H}}$ [Mpc/h]} & Reference \\
		\hline
		&         NGC        &        SGC         &  \\
		\cline{2-3}
		0.457    & $ 64.20 \pm 1.30 $ & $ 66.70 \pm 1.60 $ & \multirow{5}{*}{\makecell{\cite{Ntelis17}}}  \\
		0.511    & $ 65.40 \pm 0.90 $ & $ 63.90 \pm 1.50 $ &  \\
		0.565    & $ 62.60 \pm 0.80 $ & $ 65.20 \pm 1.60 $ &  \\
		0.619    & $ 60.40 \pm 0.80 $ & $ 60.10 \pm 1.10 $ &  \\
		0.673    & $ 59.00 \pm 0.80 $ & $ 60.10 \pm 1.80 $ &  \\
		\hline
		&         PH         &        LS          &  \\
		\cline{2-3}
		0.985    & $ 48.78 \pm 3.82 $ & $ 52.93 \pm 7.55 $ & \multirow{4}{*}{\makecell{\cite{Rodrigo18b}}} \\
		1.350    & $ 40.56 \pm 3.39 $ & $ 40.43 \pm 5.64 $ &  \\
		1.690    & $ 36.19 \pm 3.45 $ & $ 36.66 \pm 4.80 $ &  \\
		2.075    & $ 27.91 \pm 3.91 $ & $ 29.94 \pm 3.35 $ &  \\
		\hline
	\end{tabular}
\end{table}

\begin{figure*}
	\centering
	\includegraphics[scale=0.5]{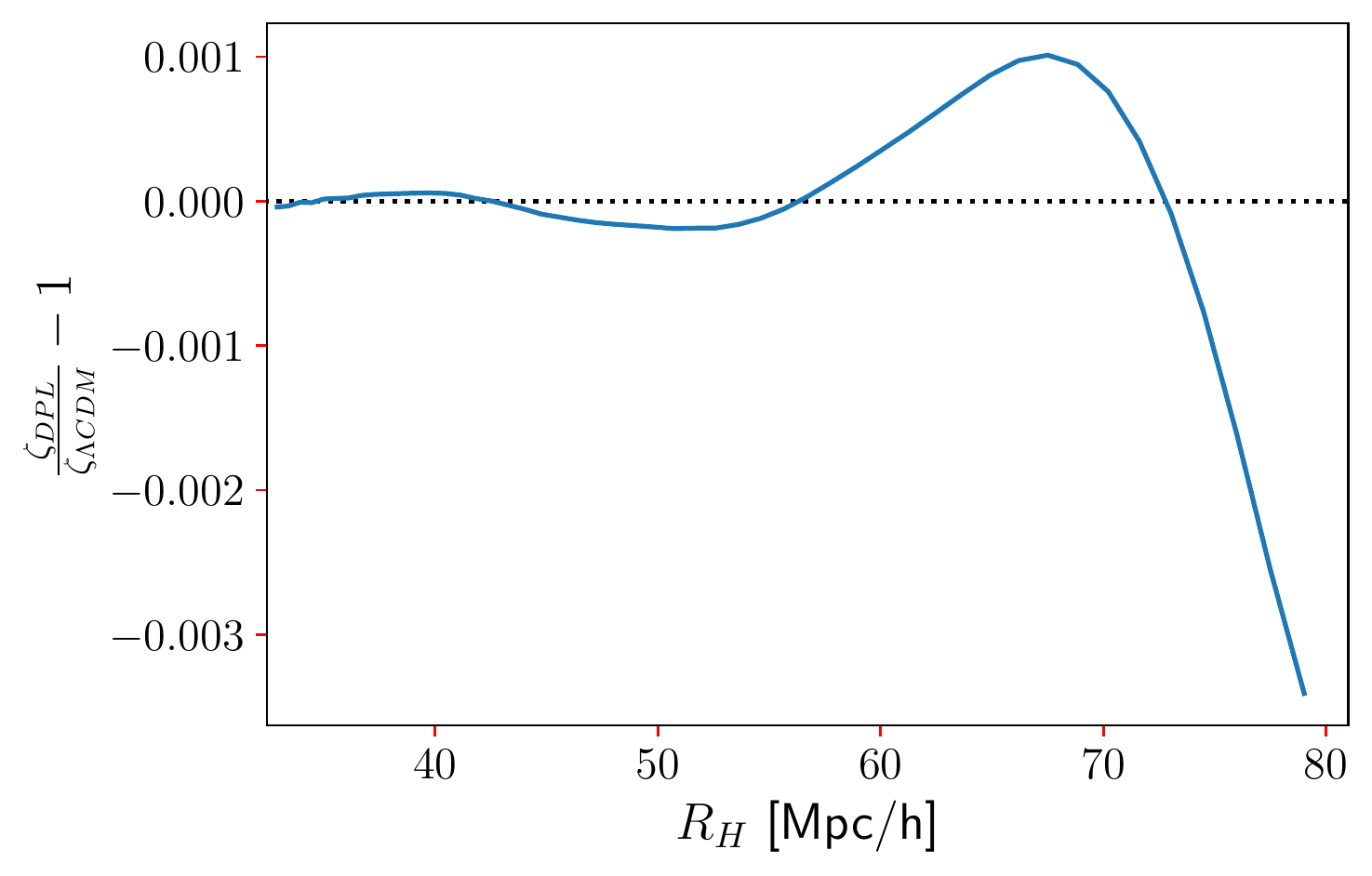} \,\,\,\,\,\,\,\,
	\includegraphics[scale=0.5]{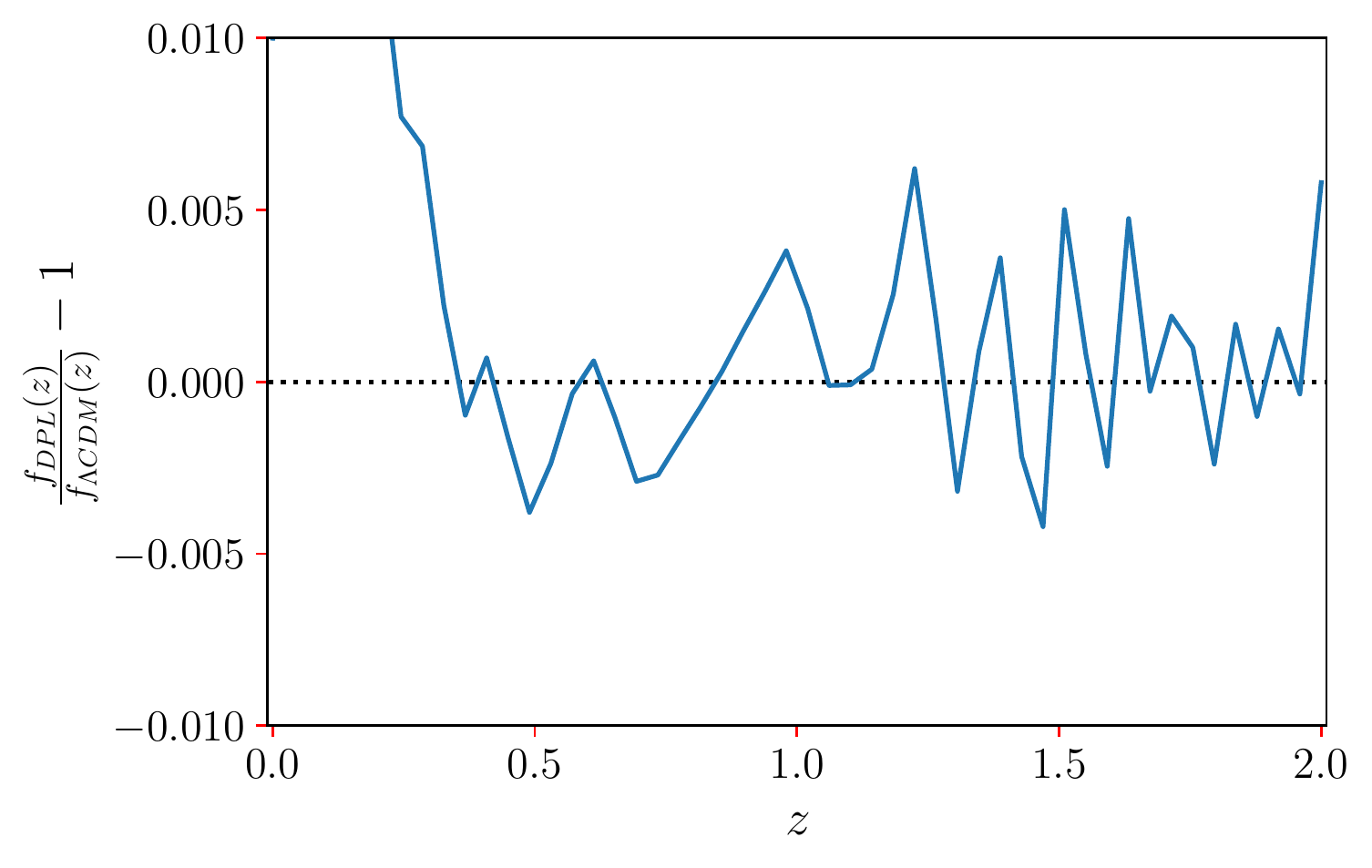}
	\caption{Left panel: Relative error between $\zeta[R_{\text{H}}]$ as fitted using the DPL approximation (equation (\ref{chiapproximation})) and the expectation from the $\Lambda$CDM model.
	Right panel: Relative error of the cosmic growth rate $f(z)$ calculated using equation (\ref{frDPLmodel}) with respect to $f(z)$ obtained using  the \textcolor{blue}{equation (\ref{growthrate})} 
	assuming the same $\Lambda$CDM fiducial cosmology; one also observes the noisy pattern caused by the numerical derivative.
	}
	\label{fig:relativeerror}
\end{figure*}

\subsection{Double Power-Law approximation for $\zeta(R_{\text{H}})$}\label{DPL}

In order to use the selected $R_{\text{H}}$ data sample to 
calculate $f^{R_{\text{H}}}(z)=f(z)$, using equation (\ref{fz_Rh}), we need to define 
$\zeta(R_{\text{H}})$. 
For this we approximate $\zeta(R_{\text{H}})$ by a Double Power Law (DPL) function, similar to 
that one used in the study of the AGN luminosity function~\citep{Kulkarni}
\begin{equation}\label{chiapproximation}
\zeta(R_{\text{H}}) =- \frac{CR_{\text{H}}^{-1}}{(R_{\text{H}}/R_{\star})^{\alpha} 
+ (R_{\text{H}}/R_{\star})^{\beta}} \, ,
\end{equation}
where $C$, $R_{\star}$, $\alpha$, and $\beta$ are the parameters to be adjusted. 
Taking its redshift derivative
\begin{equation}\label{chiterm}
\frac{1}{\zeta}\frac{d\zeta}{dz} = - \dfrac{(1 + \alpha)(R_{\text{H}}/R_{\star})^{\alpha} + (1 + \beta)(R_{\text{H}}/R_{\star})^{\beta}}{(R_{\text{H}}/R_{\star})^{\alpha} + (R_{\text{H}}/R_{\star})^{\beta}}\frac{1}{R_{\text{H}}}\frac{dR_{\text{H}}}{dz} \, ,
\end{equation}
the growth rate can be written as 
\begin{equation}\label{frDPLmodel}
f^{R_{\text{H}}}(z) = \frac{1+z}{2} 
\!\left[1 - \dfrac{(1 + \alpha) \left( \frac{R_{\text{H}}}{R_{\star}} \right)^{\alpha} 
+ (1 + \beta) \left( \frac{R_{\text{H}}}{R_{\star}} \right)^{\beta}}
{\left( \frac{R_{\text{H}}}{R_{\star}} \right)^{\alpha} 
+ \left( \frac{R_{\text{H}}}{R_{\star}} \right)^{\beta}} 
\right]
\!\frac{1}{R_{\text{H}}}\frac{dR_{\text{H}}}{dz} \, ,
\end{equation}
where $R_{\text{H}} = R_{\text{H}}(z)$. 

We fit the 4 free parameters of the DPL approximation to the theoretical expectation for $\zeta(R_{\text{H}})$, for $R_{\text{H}}$ corresponding the redshift range $0 < z < 2$, considering 
the $\Lambda$CDM model baseline obtained from~\cite{Planck18}, that is, $h=0.6727$, $\Omega_{c}h^{2}=0.1202$, $\Omega_{b}h^{2}=0.02236$, $\Sigma m_{\nu} = 0.0600$, $n_{s} = 0.9649$, $\sigma_{8}=0.8120$, and $\ln(10^{10}A_{s})=3.045$. 
For this we employ the public code \texttt{cosmopit}\footnote{\url{https://github.com/lontelis/cosmopit}}~\citep{Ntelis17,Ntelis18} to produce 
$\zeta(R_{\text{H}})^{\Lambda\text{CDM}}$, which uses  the public code \texttt{CLASS}\footnote{\url{https://github.com/lesgourg/class_public}}~\citep{Lesgourgues11a, Lesgourgues11b} as a background. 
We obtain for these parameters $[R_{\star}, \alpha, \beta, C]=[46.16, 2.76, 1.12, 0.19]$,
whose error for each of them 
is less than $1\%$. 
The plot on the left panel of Fig. \ref{fig:relativeerror} shows the relative error between the 
input $\Lambda$CDM expectation and the fitted DPL approximation, where we observe a good agreement on all scales. 
The maximum discrepancy of 0.3\% appears at the largest $R_H$ scales considered here. This occurs due to the effect introduced in $\zeta(R_H)$ by the presence of the BAO feature at $\sim 100 h^{-1}$Mpc in the correlation function, then the DPL approximation fails to model the large scales. See appendix \ref{appendixA} for more details.
Notice that, although the methodology does not depend on $\epsilon$ explicitly, as shown in equation (\ref{fz_Rh}), we follow \cite{Scrimgeour} and fix this value to $\epsilon=0.01$ to obtain the $\zeta[R_{\text{H}}(z)]^{\Lambda\text{CDM}}$ 
function and then calculate the best-fit parameters for the DPL approximation. In appendix \ref{appendixB} we test the criterion for $\epsilon$ and the dependence of our 
methodology on some  cosmological parameters.
Additionally, the right panel of Fig.  \ref{fig:relativeerror} shows the relative error between the cosmic 
growth rate $f^{R_{\text{H}}}(z)$ calculated using the DPL approximation equation (\ref{frDPLmodel}) and that obtained from 
equation~(\ref{growthrate}) assuming the $\Lambda$CDM fiducial model. 
As observed, for almost the whole redshift interval, we observe a maximum deviation of 
$\sim 0.5\%$, again at low redshifts, where we also notice the noisy pattern caused by the numerical derivative. 
Note that, since we have a small set of $R_{\text{H}}$ data, we use a reconstructed function from them to be able to appropriately 
calculate the derivative $dR_{\text{H}}/dz$ in equation \ref{frDPLmodel}, a procedure detailed in the following section.

\begin{figure*}
\centering
\includegraphics[scale=0.6]{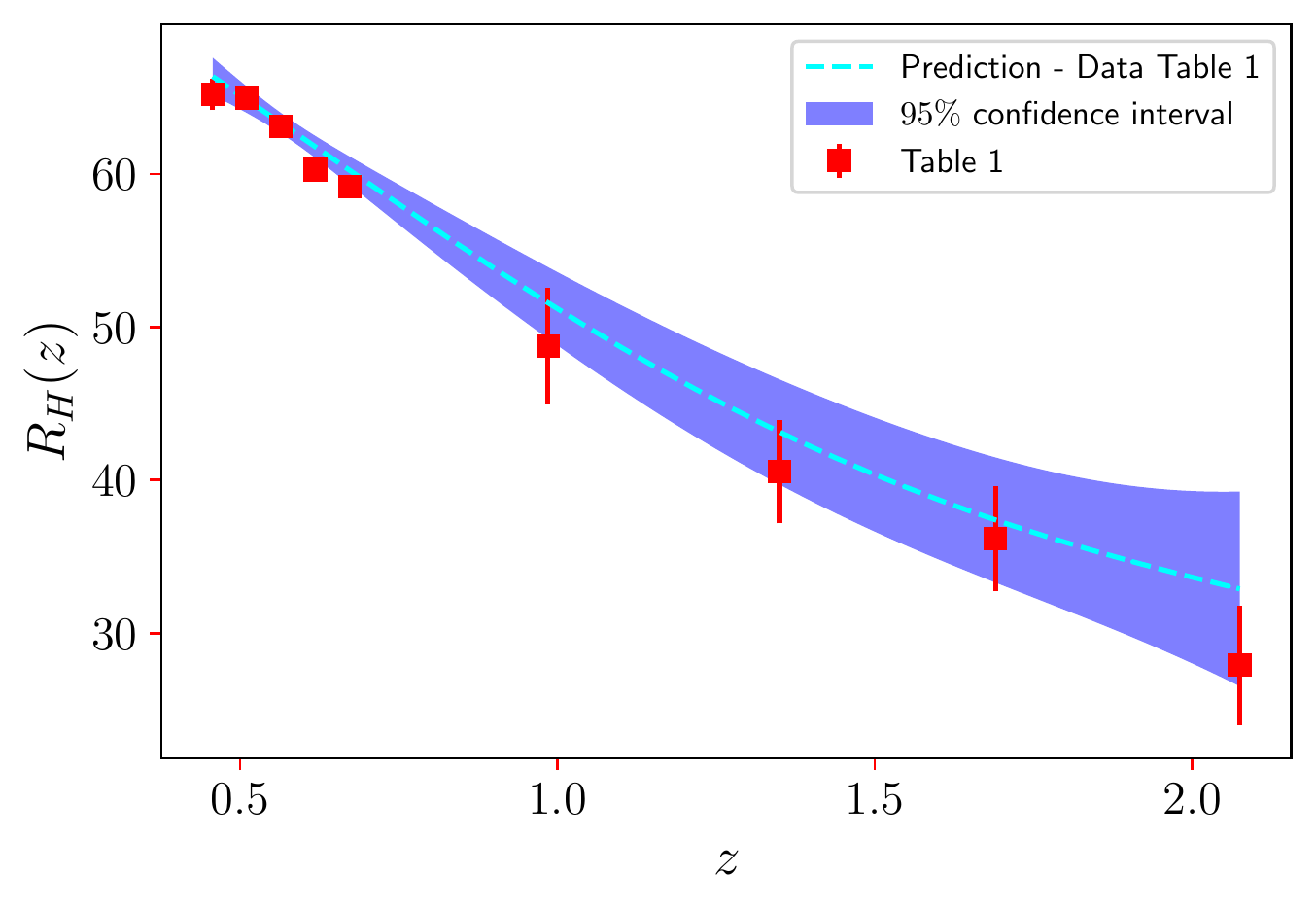} \,\,\,
\includegraphics[scale=0.6]{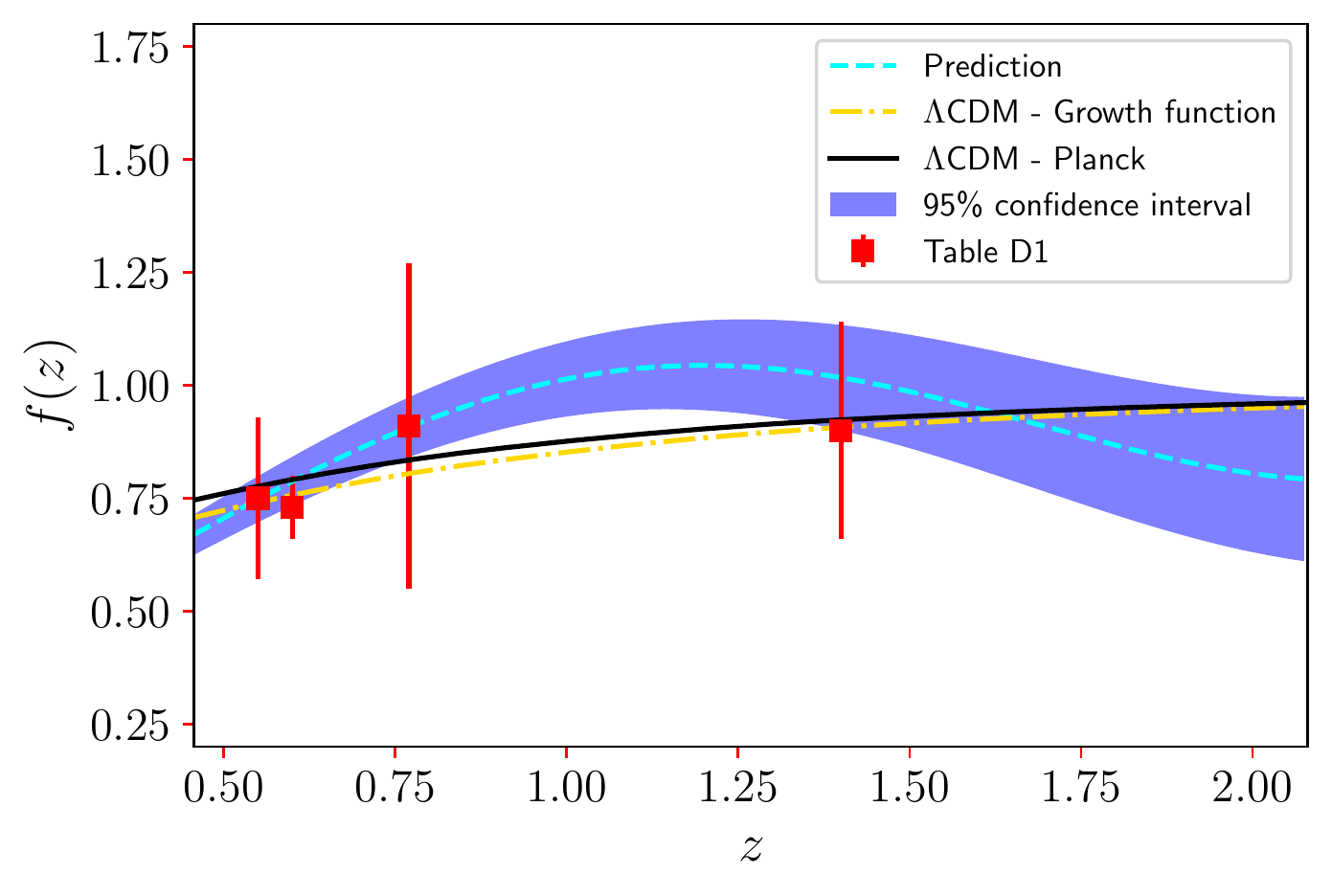} 
\caption{Left panel:  
Reconstruction of the homogeneity scale function $R_{\text{H}}(z)$ using Gaussian 
Process (dashed line) and the $R_{\text{H}}$ measurements (red squares) presented in 
Table \ref{tab:Rhdatapoints} (see the text for details about the dataset); the shadow represents the 95\% CL. 
Right panel: 
Derivation of the growth rate of structures $f^{R_{\text{H}}}(z)$ (dashed line) using the 
reconstructed 
function $R_{\text{H}}(z)$, shown in the left panel plot, and the equation (\ref{frDPLmodel}), 
where the shadow represents the 95\% CL. 
The solid line represents the expression expected in the $\Lambda$CDM model, 
$f^{\Lambda\text{CDM}}(z)$, obtained from equation (\ref{growthrate}) 
with $\Omega_m^{\text{Planck}} = 0.315$; instead, the dot-dashed line shows the 
expected growth rate using equation~(\ref{growthrate}) but with $\Omega_m = 0.270^{+0.079}_{-0.073}$, a 
value obtained from the best-fit analyses of the $f(z)$ data as shown in Fig.~\ref{fig:f} (the red squares are the data listed  in  Table~\ref{tableD1}). 
Comparing the $f^{R_{\text{H}}}(z)$ (dashed line) and the best-fit $f(z)$ (dot-dashed line) functions 
we found an agreement of $< \!2 \sigma$
(considering the corresponding uncertainties, not shown in  the figure to avoid excess of information).
}
\label{fig:frngc}
\end{figure*}

\subsection{Gaussian Process Regression}

To extract maximum cosmological information from the $R_{\text{H}}$ data listed in Table 
\ref{tab:Rhdatapoints} we perform a Gaussian Process (GP) Regression method, obtaining 
in this way a smooth curve for $R_{\text{H}}(z)$ and, by numerical derivation, for 
$dR_{\text{H}}/dz$; this information is then used in equation (\ref{frDPLmodel}) to obtain 
the $f^{R_{\text{H}}}(z)$ function. 
The GP consists of generic supervised learning method designed to solve regression and 
probabilistic classification problems, where we can interpolate the observations and 
compute empirical confidence intervals and a prediction in some region of 
interest~\citep{Rasmussen,Pezotta}. 
The GP method design from  machine learning techniques is the state-of-the-art to obtain 
statistical information and model prediction from some previously known information or 
data. 
In the cosmological context, GP techniques has been used to reconstruct cosmological 
parameters, like the dark energy equation of state, $\omega(z)$, the expansion rate of the 
universe, the cosmic growth rate, and other cosmological functions 
(see, e.g.,~\cite{Seikel12,Shafieloo12,Zhang,Marques19,Marques20,Renzi20,Nunes1,Nunes2,%
Bonilla21a,Bonilla21b,Colgain21, Sun21,Escamilla21} for a short list of references).

The main advantage in this procedure is that it is able to make a non-parametric inference using only a few physical considerations and minimal cosmological assumptions. Our aim is to reconstruct a function $F(x_i)$ from a set of its measured values $F(x_i) \pm \sigma_i$, where $x_i$ represent our data sample. It assumes that the value of the function at any point $x_i$ follows a Gaussian distribution. The value of the function at $x_i$ is correlated with the value at other point $x_i'$. Thus, a GP is defined as
\begin{equation}
\label{eqn:GPs}
F(x_i)=\mathcal{GP}(\mu(x_i),\textrm{cov}[F(x_i),F(x_i)]),
\end{equation}
where $\mu(x_i)$ and $\textrm{cov}[F(x_i),F(x_i)]$ are the mean and the variance of the variable at $x_i$, respectively. For the reconstruction of the function $F(x_i)$, the covariance between the values of this function at different positions $x_i$ can be modeled as
\begin{equation}
\label{eqn:cov}
\textrm{cov}[F(x),F(x')] = k(x,x'),
\end{equation}
where $k(x,x')$ is known as the kernel function. The kernel choice is often crucial for obtaining good results regarding the reconstruction of the function of interest.

The kernel most commonly used is the standard Gaussian Squared-Exponential approach, which is defined as
\begin{equation}
\label{eqn:kSE}
k(x,x') = \sigma_F^2 \exp\left(-\frac{|x-x'|^2}{2 l^2}\right),
\end{equation}
where $\sigma_{F}^2$ is the signal variance, which controls the strength of the correlation of the function $F$, and $l$ is the length scale that determines the capacity to model the main characteristics (global and local) in the evaluation region to be predicted (or coherence length of the correlation in $x$). These two parameters are often called hyper-parameters.

It is well known that depending on the data set in analysis, 
the kernel choice is an important point. We verify that our data set is well modeled by the choice above, and that other kernels do not produce major changes in our main results (see appendix B). In what follows, we discuss our results.

\section{Results and Discussions}\label{resulsanddiscussions}

In this section, we present our main results. We performed a GP to reconstruct the homogeneity scale, $R_H(z)$. From this, using the DPL model, we can obtain $f^{R_{\text{H}}}(z)$. Also, we perform the GP to our 
$f(z)$ data compilation. Finally, we study the parameter space $H_{0}$ -- $\Omega_{m}$ from the same data compilation.

\subsection{Results of the reconstruction of $f^{R_{\text{H}}}(z)$ and $f(z)$}\label{subsection5.1}

In obtaining the results to be described in this section, we use the Scikit-learn code 
\citep{Pedregosa}, which is a Python module integrating a wide range of state-of-the-art 
machine learning algorithms, to model the GP described in the previous section. 
The hyper-parameters $\sigma_{F}^2$ and $l$ are optimized during the fitting by maximizing 
the log-marginal-likelihood.

The left panel of Fig. \ref{fig:frngc} shows the best-fit prediction of the reconstruction GP 
of the homogeneity scale function $R_H(z)$, at 95\% CL, from the data sample listed in 
Table \ref{tab:Rhdatapoints} (represented by red squares in this plot). 
As verified in this plot, the $R_H(z)$ data reveals the expected behavior in the evolution 
of the matter clustering in the Universe, going from a nearly homogeneous situation at high
redshift to a non-linear clustered matter at low redshift where the homogeneity scale is 
attained only at large scales.

We use these $R_H(z)$ data to obtain the evolution of the growth rate of cosmic 
structures, $f^{R_{\text{H}}}(z)$, according to equation (\ref{frDPLmodel}) following 
the procedure described in section~\ref{3.1}. 
Our result can be observed on the right panel of Fig.~\ref{fig:frngc}, where 
$f^{R_{\text{H}}}(z)$ is plotted as a dashed line and the current measurements of $f(z)$, 
listed in the Table \ref{tableD1} in appendix \ref{appendixD}, as red squares. 
It is important to mention that $f^{R_{\text{H}}}$ obtained through our procedure does not 
represent a direct $f(z)$ measurement, but a non-parametric inference that can describe 
the evolution of the growth rate function from minimal cosmological assumptions. 
We also show for comparison the $\Lambda$CDM expected growth rate using 
equation~(\ref{growthrate}) in two cases: using $\Omega_m^{\text{Planck}} = 0.315$  
(continuous line) from the Planck cosmological parameters, and using $\Omega_m = 0.26$ 
(dot-dashed  line) from the best-fit data analyses shown in Fig.~\ref{fig:f}. 

\begin{figure}
\centering
\includegraphics[scale=0.6]{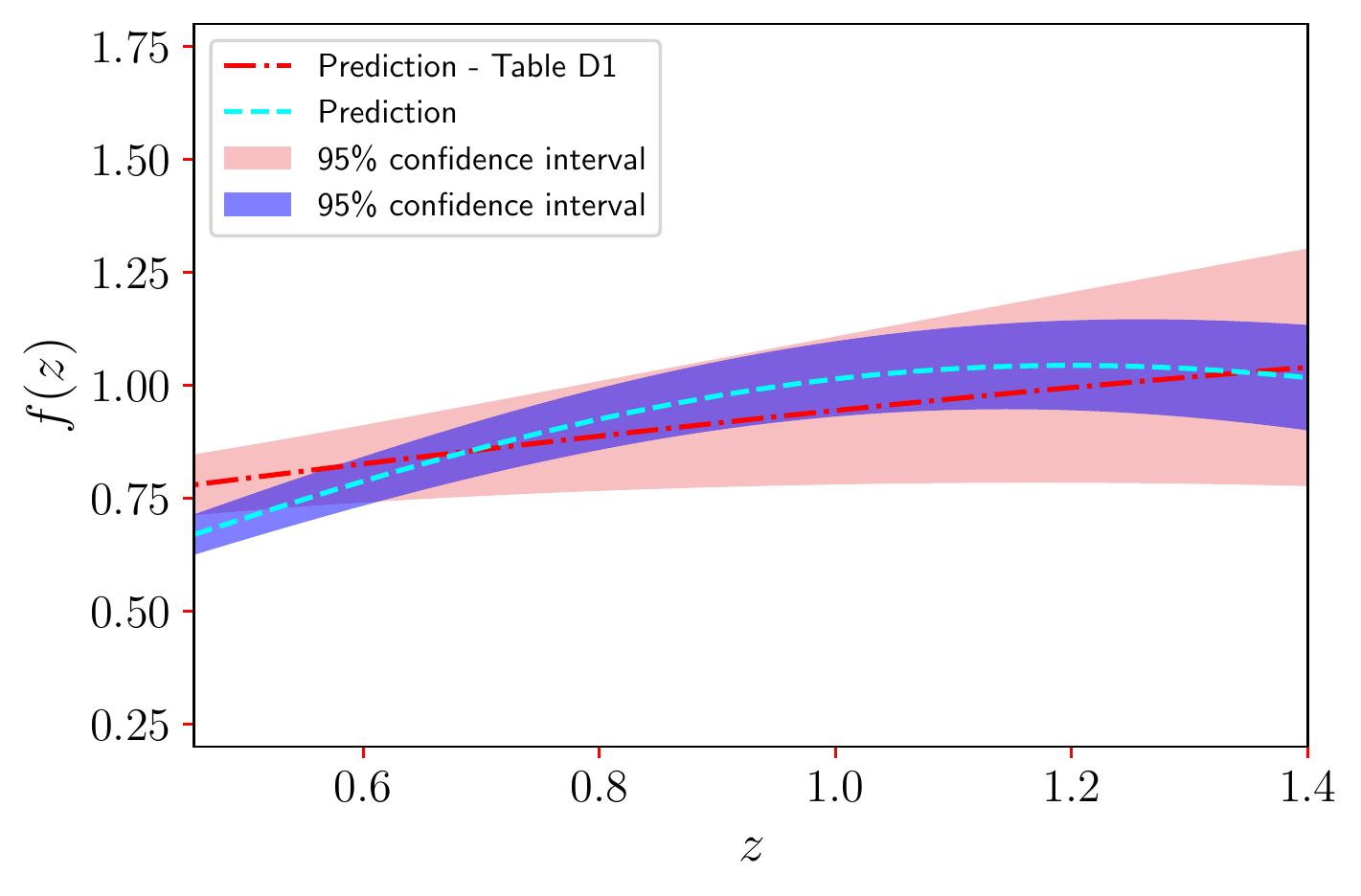}
\caption{Comparison of the growth rate $f^{R_{\text{H}}}(z)$ obtained from equation (\ref{frDPLmodel}) (dashed line) and the GP 
reconstructed function $f^{\text{data}}(z)$ (dot-dashed line) using the data compilation given in Table~\ref{tableD1}. 
Both reconstructed functions show a significant  overlapping of the respective $2 \sigma$ regions (shaded areas).
}
\label{fig:frngc_2}
\end{figure}

On the other hand, it is interesting to compare the growth rate of cosmic structures 
$f^{R_{\text{H}}}(z)$ from the evolution of the cosmic homogeneity scale, with the 
$f^{\text{data}}(z)$ resulting from a GP reconstruction using the current $f(z)$ data listed 
in Table \ref{tableD1}. 
Notice that, the reconstruction procedure of $f^{R_{\text{H}}}(z)$ is performed in the redshift interval with $R_\text{H}(z)$ data, namely $z \in [0.457,2.075]$, 
while the reconstruction procedure of $f^{\text{data}}(z)$ is done with 
$f(z)$ data in the interval $z \in [0.013, 1.4]$. 
Then, for the comparative analysis we consider
the common  redshift  interval: $z \in [0.457, 1.4]$ shown in Fig.~\ref{fig:frngc_2}, where we observe that both functions agree well ($< 2 \sigma$ level), overlapping significantly. 

One should notice that a plausible systematic present in the $R_{\text{H}}$ data, listed 
in Table~\ref{tab:Rhdatapoints}, is sourced by the necessity to assume a fiducial 
cosmology 
to calculate the 3D distances to the cosmic objects (galaxies or quasars), so that one can determine the 3D separation distance between each pair of them, information used to measure $R_{\text{H}}$. 
As a matter of fact, the $R_{\text{H}}(z)$ measurements are model dependent and one 
should be cautious with this. 
For instance, the analyses done by~\cite{Ntelis17} assumed a fiducial cosmology different 
to that assumed by~\cite{Rodrigo18b}, a fact that helps to explain why in the left panel of 
Fig.~\ref{fig:frngc} one data set appear slightly over and the other slightly under the 
reconstructed function (dashed line).

\subsection{Validation test of $H_{0}$ - $\Omega_{m}$ plane estimates from the current compilation of growth rate  data}\label{subsection5.2}

As a final discussion of this section, we will check what our compilation of $f(z)$ data, shown in Table~\ref{tableD1}, can tell us about the $\Lambda$CDM baseline. Let us perform an analysis in three steps: 

\noindent
i) To constrain $\Omega_{m}$ we consider $f(z)$ data (see Table~\ref{tableD1}) only. 

\noindent
ii) A combination $f(z)$ data plus a Gaussian prior on $H_0$ using the Planck-CMB best 
fit. Note that we are within a $\Lambda$CDM baseline, so to use Planck-CMB information 
in $\Lambda$CDM itself context is just to improve the constraint on $\Omega_{m}$. 
 
\noindent
iii) We consider the joint analysis $f(z)$ + BAO. In this work, we consider the most recent 
BAO data compilation comprised of the $D_V(z)/r_d$, $D_M(z)/r_d$, and $D_H(z)/r_d$ measurements compiled in Table~3 in~\cite{Alam21}.

We use the Metropolis-Hastings mode in \texttt{CLASS} + \texttt{MontePython} 
code \citep{Lesgourgues11a,Lesgourgues11b,Audren13,Brinckmann19} 
to derive the constraints on cosmological parameters from the data sets described above, 
ensuring a Gelman-Rubin convergence criterion of $R - 1 < 10^{-3}$.

\begin{figure}
\centering
\includegraphics[scale=0.75]{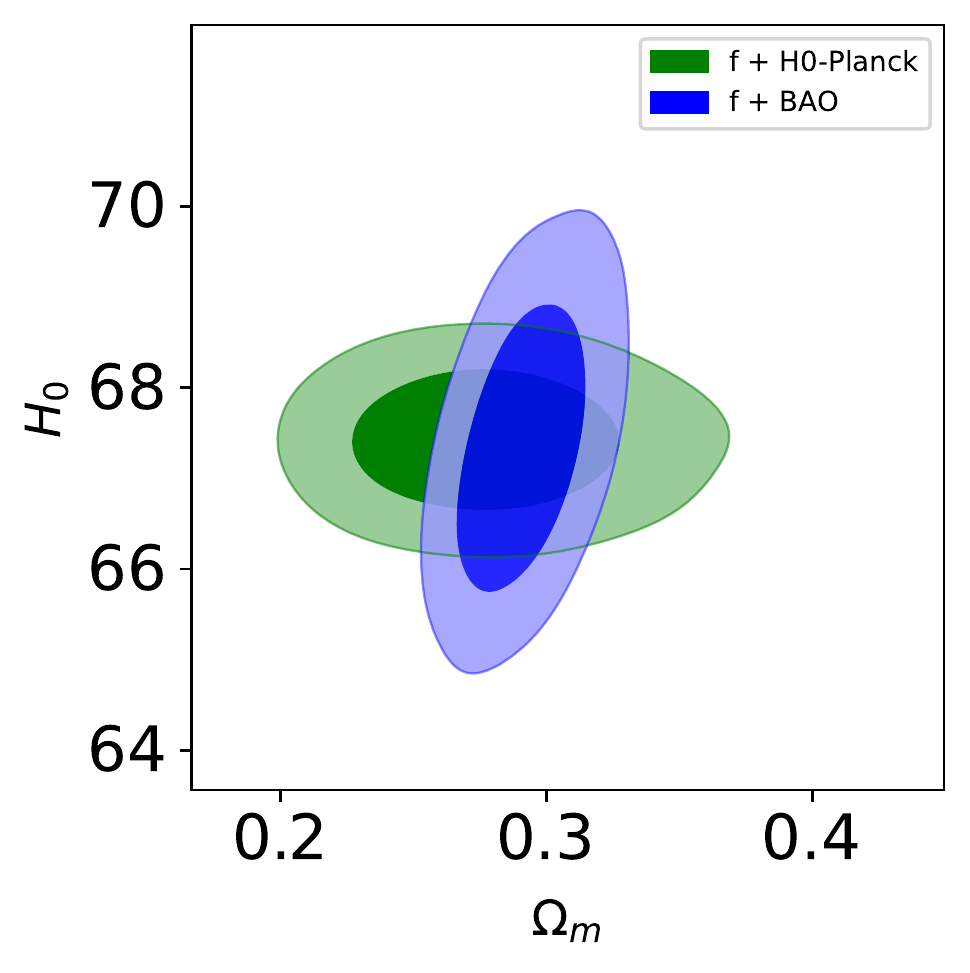}
\caption{The 68\% (dark shaded area) and 95\%  
(light shaded area) CLs regions, respectively, on the parametric space $H_0 - \Omega_m$ from $f(z)$ + Planck-$H_0$ prior and $f(z)$ + BAO joint analyses. The parameter $H_{\rm 0}$ is measured 
in the units of km s${}^{-1}$ Mpc${}^{-1}$.}
\label{fig:f}
\end{figure}

Figure \ref{fig:f} shows the parameter space in the $H_0 - \Omega_m$ plane at 68\% and 
95\% CL from $f(z)$ + Planck-$H_0$ prior and $f(z)$ + BAO joint analyses, 
where $f(z)$ data refers to the measurements presented in Table~\ref{tableD1}. 
The summary of the main results of our statistical analyses at 68\% CL are: 
$\Omega_m = 0.27^{+0.079}_{-0.073}$ ($f(z)$ data only), 
$\Omega_m = 0.279^{+0.066}_{-0.067}$ 
($f(z)$ + $H_0$-Planck) and $\Omega_m = 0.291^{+0.033}_{-0.030}$ and 
$H_0 =67.4^{+2.1}_{-2.0}$ km s${}^{-1}$ Mpc${}^{-1}$ from $f(z)$ + BAO combination.

As well known, there is a growing tension for low $z$ measurements of $f(z)$ and it is weaker than the Planck-$\Lambda$CDM predictions (see \cite{DiValentino21,Perivolaropoulos21} and reference therein for a review and the recent discussion presented in \cite{Nunes21}). Our results here also confirm that growth rate data based on the measurements in Table \ref{tableD1} predict a suppression on the amplitude of the matter density perturbation at low $z$ due the low $\Omega_m$ estimation in comparison with that from the Planck-$\Lambda$CDM, $\Omega_m = 0.315 \pm 0.007$ \citep{Planck18}. 
On the right panel of Fig. \ref{fig:frngc} we also show the theoretical curve assuming our 
constraint on $\Omega_m$. 
Despite predicting a low $\Omega_m$ best fit value in our analysis, the error bar estimates are in agreement with Planck-CMB at $< 1\sigma$.

\section{Conclusions}\label{conclusions}

Measurements of the growth rate of cosmic structures, $f$, have the potential to 
differentiate between the theory of general relativity, that supports the concordance model 
$\Lambda$CDM, from alternative scenarios based on modified gravity models. 
Besides the efforts, the current uncertainties in such measurements do not allow to 
discern between competing models of modified gravity. 

This motivated us to search for a cosmological observable that depends only on the 
cosmic time $t$, or equivalently on the redshift $z$. 
We propose to use the transition scale to homogeneity, $R_{\text{H}}(z)$, to know the 
evolution of the growth rate $f=f(z)$. 
As shown in the section~\ref{3.1}, the relation between $R_{\text{H}}$ and $f$ is not 
direct and one needs two ingredients: 
(i) a set of $\{ R_{\text{H}}(z_i) \}$ data --in the redshift interval of interest-- to reconstruct 
the continuous function $R_{\text{H}}(z)$ and to perform its redshift derivative; and 
(ii) the matter two-point correlation function at $z=0$, $\xi(r, z=0)$, that analyzes 
distance scales of the order of the homogeneity scale. 
However, there is no observational data to construct $\xi(r, z=0)$, and one has to assume 
a fiducial cosmology to obtain it. 
For this reason, our analyses and results are actually consistency tests of the cosmological 
model assumed. 

Using GP, our reconstruction of the homogeneity scale function $R_{\text{H}}(z)$ done in 
section~\ref{resulsanddiscussions} shows the expected behavior, although the current 
dataset is small and with large errors (see Table~\ref{tab:Rhdatapoints}). 
With the functions $R_{\text{H}}(z)$ and $\xi(r, z=0)$, and following our procedure, we 
use them in equation~(\ref{frDPLmodel}) to obtain the growth rate of cosmic structures 
$f^{R_{\text{H}}}(z)$. 
Our results, displayed in the right panel of fig.~\ref{fig:frngc}, show a good agreement 
between: 
(i) the growth rate $f^{R_{\text{H}}}(z)$ obtained through our approach, 
(ii) the $f^{\Lambda\text{CDM}}(z)$ expected in the fiducial model, and 
(iii) the best-fit $f(z)$ from the set of $\{ f(z_i) \}$ measurements available in the literature. 
Moreover, using this compilation of $\{ f(z_i) \}$ data (see Appendix~\ref{appendixD}), we 
perform a GP to reconstruct the growth rate function $f^{\text{data}}(z)$ 
and compare it with the function $f^{R_{\text{H}}}(z)$ obtained from our approach, 
finding a concordance of $< \!2 \sigma$ as observed in Fig.~\ref{fig:frngc_2} 
(notice the significant overlapping of their  $2 \sigma$ regions).
This is a good result considering the few data available for both reconstruction processes.

It is worth to note that our approach to find the growth rate of cosmic structures, $f(z)$, 
from the evolution of the cosmic homogeneity scale, $R_{\text{H}}(z)$, 
relies on the definition of the homogeneity scale which is not unique (see, e.g.,~\cite{Pandey21a,Pandey21b}); 
in our approach the homogeneity scale is provided by the estimator $D_2$ through the 
analyses of the universe fractal structure~\citep{Scrimgeour}.

The relationship found between $f$ and $R_{\text{H}}$ indicates that with precise 
homogeneity scale data, $R_{\text{H}}(z)$, measured at several redshifts from forthcoming surveys~(see,  e.g.,~\cite{Amendola18,Ivezic19}), one can determine with good accuracy the growth rate 
of cosmic structures $f=f(z)$, which in turn can be used to discriminate between the 
concordance $\Lambda$CDM and competing models based on  modified gravity theories. 
Moreover, these data could be used in statistical analyses to find cosmological parameters. 
In summary, the homogeneity scale data, $R_{\text{H}}(z)$, would indeed play the role of a novel cosmological  observable, as first discussed by~\citet{Ntelis19}.


\section*{Acknowledgements}
FA, AB, EdC, and CPN thank CAPES, CNPq, PROPG-CAPES/FAPEAM program, 
and FAPESP (process no. 2019/06040-0) for the grants under which this work was 
carried out. 
RCN acknowledges financial support from the Funda\c{c}\~{a}o de Amparo \`{a} Pesquisa 
do Estado de S\~{a}o Paulo (FAPESP, S\~{a}o Paulo Research Foundation) under the project no. 2018/18036-5.

\section*{Data Availability}
The data underlying this article will be shared on request to the corresponding author.




\begin{thebibliography}{99}





\bibitem[\protect\citeauthoryear{Alam et al.}{2020}]{Alam20}
Alam, S., et al., 2020, 
[\href{https://arxiv.org/abs/2011.05771.pdf}{arXiv:2011.05771}]

\bibitem[\protect\citeauthoryear{Alam et al.}{2021}]{Alam21}
Alam, S., et al., 2021, \href{https://journals.aps.org/prd/abstract/10.1103/PhysRevD.103.083533}{Phys. Rev. D}, 103, 083533, [\href{https://arxiv.org/abs/2007.08991}{arXiv:2007.08991}]

\bibitem[\protect\citeauthoryear{Alonso et al.}{2015}]{Alonso15}
Alonso, D., et al., 2015, \href{https://academic.oup.com/mnras/article/449/1/670/1321149}{MNRAS}, 449, 670, [\href{https://arxiv.org/abs/1412.5151}{arXiv:1412.5151}]

\bibitem[\protect\citeauthoryear{Amendola et al.}{2018}]{Amendola18}
Amendola, L., et al., 2018, \href{https://link.springer.com/article/10.1007/s41114-017-0010-3}{Living Rev. Rel.}, 21, 2, [\href{https://arxiv.org/abs/1606.00180}{arXiv:1606.00180}]


\bibitem[\protect\citeauthoryear{Aubert et al.}{2020}]{Aubert20}
Aubert, M., et al., 2020,  [\href{https://arxiv.org/abs/2007.09013}{arXiv:2007.09013}]

\bibitem[\protect\citeauthoryear{Audren et al.}{2013}]{Audren13}
Audren, B., et al., 2013, \href{https://iopscience.iop.org/article/10.1088/1475-7516/2013/02/001}{J. Cosmol. Astropart. Phys.}, 02, 001, [\href{https://arxiv.org/abs/1210.7183}{ arXiv:1210.7183}]

\bibitem[\protect\citeauthoryear{Avila et al.}{2018}]{Avila18}
Avila, F., et al., 2018, \href{https://iopscience.iop.org/article/10.1088/1475-7516/2018/12/041}{J. Cosmol. Astropart. Phys.}, 12, 041, [\href{https://arxiv.org/abs/1806.04541}{arXiv:1806.04541}]

\bibitem[\protect\citeauthoryear{Avila et al.}{2019}]{Avila19}
Avila, F., et al., 2019, \href{https://academic.oup.com/mnras/article-abstract/488/1/1481/5526236?redirectedFrom=fulltext}{MNRAS}, 488, 1481, [\href{https://arxiv.org/abs/1906.10744}{arXiv:1906.10744}]

\bibitem[\protect\citeauthoryear{Avila et al.}{2021}]{Avila21}
Avila, F., et al., 2021, \href{https://academic.oup.com/mnras/article-abstract/505/3/3404/6284050?redirectedFrom=fulltext}{MNRAS}, 505, 3404, [\href{https://arxiv.org/abs/2105.10583}{arXiv:2105.10583}]


\bibitem[\protect\citeauthoryear{Basilakos}{2012}]{Basilakos12}
Basilakos, S., 2012, \href{https://www.worldscientific.com/doi/abs/10.1142/S0218271812500642}{Int. J. Mod. Phys. D}, 21, 1250064, [\href{https://arxiv.org/abs/1202.1637}{arXiv:1202.1637}]

\bibitem[\protect\citeauthoryear{Basilakos, \& Anagnostopoulos}{2020}]{Basilakos20}
Basilakos, S., Anagnostopoulos, F. K., 2020, \href{https://link.springer.com/article/10.1140/epjc/s10052-020-7770-8}{Eur. Phys. J. C}, 80, 1, [\href{https://arxiv.org/abs/1903.10758}{arXiv:1903.10758}]

\bibitem[\protect\citeauthoryear{Bautista et al.}{2021}]{Bautista21}
Bautista, J. E., et al., 2021, \href{https://academic.oup.com/mnras/article-abstract/500/1/736/5907692?redirectedFrom=fulltext}{MNRAS}, 500, 736, [\href{URL}{arXiv:2007.08993}]

\bibitem[\protect\citeauthoryear{Bengaly et al.}{2017}]{Bengaly17}
Bengaly, C. A. P., Bernui, A., Ferreira, I. S., Alcaniz, J. S., 2017, 
\href{https://academic.oup.com/mnras/article/466/3/2799/2679636}{MNRAS}, 466, 2799, 
[\href{https://arxiv.org/abs/1511.09414}{arXiv:1511.09414}]

\bibitem[\protect\citeauthoryear{Bernui et al.}{2007}]{Bernui07}
Bernui, A., Mota, B., Rebou\c{c}as, M. J., Tavakol, R., 2007, 
\href{https://www.worldscientific.com/doi/abs/10.1142/S0218271807010195}
{Int. J. Mod. Phys. D}, 16, 411, 
[\href{https://arxiv.org/abs/0706.0575}{arXiv:0706.0575}]

\bibitem[\protect\citeauthoryear{Bernui, Oliveira, \& Pereira}{2014}]{Bernui14}
Bernui, A., Oliveira, A. F., Pereira, T. S., 2014, 
\href{https://iopscience.iop.org/article/10.1088/1475-7516/2014/10/041}
{J. Cosmol. Astropart. Phys.}, 10, 041, 
[\href{https://arxiv.org/abs/1404.2936}{arXiv:1404.2936}]


\bibitem[\protect\citeauthoryear{Blake et al.}{2011}]{Blake11}
Blake, C., et al., 2011, \href{https://academic.oup.com/mnras/article/415/3/2876/1053484}{MNRAS}, 415, 2876, [\href{https://arxiv.org/abs/1104.2948}{arXiv:1104.2948}]

\bibitem[\protect\citeauthoryear{Blake et al.}{2013}]{Blake13}
Blake, C., et al., 2013, \href{https://academic.oup.com/mnras/article/436/4/3089/985191}{MNRAS}, 436, 3089, [\href{https://arxiv.org/abs/1309.5556}{arXiv:1309.5556}]

\bibitem[\protect\citeauthoryear{Blas, Lesgourgues, \& Tram}{2011}]{Lesgourgues11b}
Blas, D., Lesgourgues, J.,  Tram, T., 2011, \href{https://iopscience.iop.org/article/10.1088/1475-7516/2011/07/034}{J. Cosmol. Astropart. P.}, 07, 034, [\href{https://arxiv.org/abs/1104.2933}{arXiv:1104.2933}]

\bibitem[\protect\citeauthoryear{Brinckmann, \& Lesgourgues}{2019}]{Brinckmann19}
Brinckmann, T., Lesgourgues, J., 2019, \href{https://www.sciencedirect.com/science/article/abs/pii/S2212686418302309?via\%3Dihub}{Phys. Dark Universe}, 24, 100260, [\href{https://arxiv.org/abs/1804.07261}{arXiv:1804.07261}]

\bibitem[\protect\citeauthoryear{Bonilla, Kumar, \&  Nunes}{2021a}]{Bonilla21a}
Bonilla, A., Kumar, S., Nunes, R. C., 2021, \href{https://link.springer.com/article/10.1140\%2Fepjc\%2Fs10052-021-08925-z}{Eur. Phys. J. C}, 81, 1, [\href{https://arxiv.org/abs/2011.07140}{arXiv:2011.07140}]

\bibitem[\protect\citeauthoryear{Bonilla et al.}{2021b}]{Bonilla21b}
Bonilla, A., et al., 2021, [\href{https://arxiv.org/abs/2102.06149}{arXiv:2102.06149}]



\bibitem[\protect\citeauthoryear{Camacho \& Gazta\~{n}aga}{2021}]{Camacho21}
Camacho, B., Gazta\~{n}aga, E., 2021, [\href{https://arxiv.org/abs/2106.14303}{arXiv:2106.14303}] 

\bibitem[\protect\citeauthoryear{Carvalho et al.}{2020}]{Carvalho}
Carvalho, G. C., et al., 2020, 
\href{https://www.sciencedirect.com/science/article/abs/pii/S0927650520300050?via%3Dihub}{Astroparticle Physics}, 119, 102432, 
[\href{https://arxiv.org/abs/1709.00271}{arXiv:1709.00271}]

\bibitem[\protect\citeauthoryear{Colg\'{a}in, \& Sheikh-Jabbari}{2021}]{Colgain21}
Colg\'{a}in, E. \'{O}., Sheikh-Jabbari, M. M., 2021, [\href{https://arxiv.org/abs/2101.08565}{arXiv:2101.08565}]


\bibitem[\protect\citeauthoryear{Da \^{A}ngela et al.}{2008}]{DaAngela08}
Da \^{A}ngela, J., et al., 2008, \href{https://academic.oup.com/mnras/article/383/2/565/993314}{MNRAS}, 383, 565, [\href{https://arxiv.org/abs/astro-ph/0612401}{arXiv:astro-ph/0612401}]

\bibitem[\protect\citeauthoryear{Dainotti, Del Vecchio, \& Tarnopolski}{2018}]{Dainotti18}
 Dainotti, M. G., Del Vecchio, R., Tarnopolski, M., 2018, 
\href{https://www.hindawi.com/journals/aa/2018/4969503/}{Advances in Astronomy}, 
2018, 4969503, 
[\href{https://arxiv.org/abs/1612.00618}{arXiv:1612.00618}]

\bibitem[\protect\citeauthoryear{de Carvalho et al.}{2018}]{deCarvalho18}
de Carvalho, E., et al.,  2018, \href{https://iopscience.iop.org/article/10.1088/1475-7516/2018/04/064}
{J. Cosmol. Astropart. Phys.}, 04, 064, [\href{https://arxiv.org/abs/1709.00113}{arXiv:1709.00113}]


\bibitem[\protect\citeauthoryear{de Carvalho et al.}{2021}]{deCarvalho21}
de Carvalho, E., et al., 2021, 
\href{https://www.aanda.org/articles/aa/abs/2021/05/aa39936-20/aa39936-20.html}
{A\&A}, 649, A20,
[\href{https://arxiv.org/pdf/2103.14121}{arXiv:2103.14121}]

\bibitem[\protect\citeauthoryear{De Marzo, Labini, \& Pietronero}{2021}]{deMarzo21}
De Marzo, G., Labini, F. S., Pietronero, L., 2021, \href{https://www.aanda.org/articles/aa/abs/2021/07/aa41081-21/aa41081-21.html}{A\&A}, 651, A114, [\href{https://arxiv.org/abs/2105.06110}{arXiv:2105.06110}]

\bibitem[\protect\citeauthoryear{Di Valentino et al.}{2021}]{DiValentino21}
Di Valentino, E., et al., 2021, \href{https://www.sciencedirect.com/science/article/abs/pii/S0927650521000487?via\%3Dihub}{Astropart. Phys.}, 131, 102604, [\href{https://arxiv.org/abs/2008.11285}{arXiv:2008.11285}]




\bibitem[\protect\citeauthoryear{Eisenstein \& Hu}{1998}]{EH98}
Eisenstein, D. J., Hu, W., 1998, \href{https://iopscience.iop.org/article/10.1086/305424}{ApJ} 
496, 605, [\href{https://arxiv.org/abs/astro-ph/9709112}{arXiv:astro-ph/9709112}]

\bibitem[\protect\citeauthoryear{Escamilla-Rivera, Said, \& Mifsud}{2021}]{Escamilla21}
Escamilla-Rivera, C., Said, J. L., \& Mifsud, J. 2021, \href{https://iopscience.iop.org/article/10.1088/1475-7516/2021/10/016}{J. Cosmol. Astropart. Phys.}, 10, 016, [\href{https://arxiv.org/abs/2105.14332}{arXiv:2105.14332}]



\bibitem[\protect\citeauthoryear{Gon\c{c}alves et al.}{2018a}]{Rodrigo18a}
Gon\c{c}alves, R. S., et al., 2018a, \href{https://academic.oup.com/mnrasl/article/475/1/L20/4768253}{MNRASL}, 475, L20, [\href{https://arxiv.org/abs/1710.02496}{arXiv:1710.02496}]

\bibitem[\protect\citeauthoryear{Gon\c{c}alves et al.}{2018b}]{Rodrigo18b}
Gon\c{c}alves, R. S., et al., 2018b, \href{https://academic.oup.com/mnras/article-abstract/481/4/5270/5114592?redirectedFrom=fulltext}{MNRAS}, 481, 5270, [\href{https://arxiv.org/abs/1809.11125}{arXiv:1809.11125}]

\bibitem[\protect\citeauthoryear{Gon\c{c}alves et al.}{2021}]{Rodrigo21}
Gon\c{c}alves, R. S., 2021, \href{https://iopscience.iop.org/article/10.1088/1475-7516/2021/03/029}{J. Cosmol. Astropart. Phys.}, 03, 029, [\href{https://arxiv.org/abs/2010.06635}{arXiv:2010.06635}]

\bibitem[\protect\citeauthoryear{Guzzo et al.}{2008}]{Guzzo08} Guzzo, L., 2008, \href{https://www.nature.com/articles/nature06555}{Nature}, 451, 541, [\href{https://arxiv.org/abs/0802.1944}{arXiv:0802.1944}]


\bibitem[\protect\citeauthoryear{Hamilton}{1992}]{Hamilton92}
Hamilton, A. J. S., 1992, \href{http://articles.adsabs.harvard.edu/pdf/1992ApJ...385L...5H}{ApJ}, 385, L5 

\bibitem[\protect\citeauthoryear{Hamilton \& Culhane}{1995}]{Hamilton95}
Hamilton, A. J. S., Culhane, M., 1995, \href{http://articles.adsabs.harvard.edu/pdf/1996MNRAS.278...73H}{MNRAS}, 278, 73, [\href{https://arxiv.org/abs/astro-ph/9507021}{arXiv:astro-ph/9507021}]

\bibitem[\protect\citeauthoryear{Hawkins et al.}{2003}]{Hawkins03}
Hawkins, E., et al., 2003, \href{https://academic.oup.com/mnras/article/346/1/78/2891845}{MNRAS}, 346, 78, [\href{https://arxiv.org/abs/astro-ph/0212375}{arXiv:astro-ph/0212375}]

\bibitem[\protect\citeauthoryear{Heinesen}{2020}]{Heinesen20}
Heinesen, A., 2020, \href{https://iopscience.iop.org/article/10.1088/1475-7516/2020/10/052}{J. Cosmol. Astropart. Phys.}, 10, 052, [\href{https://arxiv.org/abs/2006.15022}{arXiv:2006.15022}]

\bibitem[\protect\citeauthoryear{Huterer et al.}{2015}]{Huterer15}
Huterer, D., et al., 2015, \href{https://www.sciencedirect.com/science/article/abs/pii/S0927650514001005?via\%3Dihub}{Astropart. Phys.}, 63, 23, [\href{https://arxiv.org/abs/1309.5385}{arXiv:1309.5385}]


\bibitem[\protect\citeauthoryear{Ivezi\'c, et al.}{2019}]{Ivezic19}
Ivezi\'c, \v{Z}., et al., 2019, \href{https://iopscience.iop.org/article/10.3847/1538-4357/ab042c}{ApJ}, 873, 111, [\href{https://arxiv.org/abs/0805.2366}{arXiv:0805.2366}]


\bibitem[\protect\citeauthoryear{Juszkiewicz et al.}{2009}]{Juszkiewicz}
Juszkiewicz, R., et al., 2009, \href{https://iopscience.iop.org/article/10.1088/1475-7516/2010/02/021}{J. Cosmol. Astropart. P.}, 2, 021, [\href{https://arxiv.org/abs/0901.0697}{arXiv:0901.0697}]


\bibitem[\protect\citeauthoryear{Kaiser}{1987}]{Kaiser87}
Kaiser, N., 1987, \href{http://articles.adsabs.harvard.edu/pdf/1987MNRAS.227....1K}{MNRAS}, 227, 1

\bibitem[\protect\citeauthoryear{Kazantzidis, \& Perivolaropoulos}{2018}]{Kasa18}
Kazantzidis, L., Perivolaropoulos, L., 2018, \href{https://journals.aps.org/prd/abstract/10.1103/PhysRevD.97.103503}{Phys. Rev. D}, 97, 103503, [\href{https://arxiv.org/abs/1803.01337}{arXiv:1803.01337}]

\bibitem[\protect\citeauthoryear{Kulkarni, Worseck \& Hennawi}{2019}]{Kulkarni}
Kulkarni, G., Worseck, G., Hennawi, J. F., 2019, \href{https://academic.oup.com/mnras/article-abstract/488/1/1035/5510422?redirectedFrom=fulltext}{MNRAS}, 488, 1035, [\href{https://arxiv.org/abs/1807.09774}{arXiv:1807.09774}]


\bibitem[\protect\citeauthoryear{Lahav et al.}{1991}]{Lahav91}
Lahav, O., et al., 1991, \href{https://academic.oup.com/mnras/article/251/1/128/1001136}{MNRAS}, 251, 128

\bibitem[\protect\citeauthoryear{Laurent et al.}{2016}]{Laurent16}
Laurent, P., 2016, \href{https://iopscience.iop.org/article/10.1088/1475-7516/2016/11/060}{J. Cosmol. Astropart. P.}, 11, 060, [\href{https://arxiv.org/abs/1602.09010}{arXiv:1602.09010}]

\bibitem[\protect\citeauthoryear{Lesgourgues}{2011}]{Lesgourgues11a}
Lesgourgues, J., 2011, [\href{https://arxiv.org/abs/1104.2932}{arXiv:1104.2932}]


\bibitem[\protect\citeauthoryear{Linder}{2005}]{Linder05}
Linder, E. V., 2005, \href{https://journals.aps.org/prd/abstract/10.1103/PhysRevD.72.043529}{Phys. R. D}, 72, 043529, [\href{https://arxiv.org/abs/astro-ph/0507263}{arXiv:astro-ph/0507263}]

\bibitem[\protect\citeauthoryear{Linder \& Cahn}{2007}]{Linder07}
Linder, E. V., Cahn, R. N., 2007, \href{https://www.sciencedirect.com/science/article/abs/pii/S0927650507001326?via\%3Dihub}{Astropart. Phys.}, 28, 481, [\href{https://arxiv.org/abs/astro-ph/0701317}{astro-ph/0701317v2}]

\bibitem[\protect\citeauthoryear{Linder}{2020}]{Linder20}
Linder, E. V., 2020, \href{https://iopscience.iop.org/article/10.1088/1475-7516/2020/10/042}
{J. Cosmol. Astropart. P.}, 10, 042, [\href{https://arxiv.org/abs/2003.10453}{arXiv:2003.10453}]


\bibitem[\protect\citeauthoryear{Marques et al.}{2018}]{Marques18}
Marques, G. A., Novaes, C. P., Bernui A., Ferreira, I. S., 2018, 
\href{https://academic.oup.com/mnras/article-abstract/473/1/165/4103556?redirectedFrom=fulltext}
{MNRAS}, 473, 165, [\href{https://arxiv.org/abs/1708.09793}{arXiv:1708.09793}]

\bibitem[\protect\citeauthoryear{Marques et al.}{2019}]{Marques19}
Marques, G. A., et al., 2019, 
\href{https://iopscience.iop.org/article/10.1088/1475-7516/2019/06/019}
{J. Cosmol. Astropart. P.}, 06, 019, [\href{https://arxiv.org/abs/1812.08206}
{arXiv:1812.08206}]

\bibitem[\protect\citeauthoryear{Marques et al.}{2020}]{Marques20}
Marques, G. A., Liu, J., Huffenberger, K. M., Colin Hill, J., 2020, 
\href{https://iopscience.iop.org/article/10.3847/1538-4357/abc003}{ApJ}, 904, 182, 
[\href{https://arxiv.org/abs/2008.04369}{arXiv:2008.04369}]


\bibitem[\protect\citeauthoryear{Mukhanov, Feldman, \& Brandenberger}{1992}]{Mukhanov}
Mukhanov, V. F., Feldman, H. A.,  Brandenberger, R. H., 1992, \href{https://www.sciencedirect.com/science/article/abs/pii/037015739290044Z?via\%3Dihub}{Phys. Rep.}, 215, 203


\bibitem[\protect\citeauthoryear{Ntelis et al.}{2017}]{Ntelis17}
Ntelis, P., et al., 2017, \href{https://iopscience.iop.org/article/10.1088/1475-7516/2017/06/019}{J. Cosmol. Astropart. P.}, 06, 019, [\href{https://arxiv.org/abs/1702.02159}{arXiv:1702.02159}]

\bibitem[\protect\citeauthoryear{Ntelis et al.}{2018}]{Ntelis18}
Ntelis, P., et al., 2018, \href{https://iopscience.iop.org/article/10.1088/1475-7516/2018/12/014}{J. Cosmol. Astropart. P.}, 12, 014, [\href{https://arxiv.org/abs/1810.09362}{arXiv:1810.09362}]

\bibitem[\protect\citeauthoryear{Ntelis et al.}{2019}]{Ntelis19}
Ntelis, P., et al., 2019, 
[\href{https://arxiv.org/abs/1904.06135.pdf}{arXiv:1904.06135}]

\bibitem[\protect\citeauthoryear{Ntelis et al.}{2020}]{Ntelis20}
Ntelis, P., et al., 2020, 
[\href{https://arxiv.org/abs/2010.06707.pdf}{arXiv:2010.06707}]

\bibitem[\protect\citeauthoryear{Nunes et al.}{2016}]{Nunes16}
Nunes, R. C., et al., 2016, \href{https://iopscience.iop.org/article/10.1088/1475-7516/2016/08/051}{J. Cosmol. Astropart. Phys.}, 08, 051, [\href{https://arxiv.org/abs/1509.05059}{arXiv:1509.05059}]

\bibitem[\protect\citeauthoryear{Nunes et al.}{2020a}]{Nunes1}
Nunes, R. C., Yadav, S. K., Jesus, J. F., Bernui, A., 2020a, 
\href{https://academic.oup.com/mnras/article-abstract/497/2/2133/5870123?redirectedFrom=fulltext}{MNRAS}, 497, 2133, 
[\href{https://arxiv.org/abs/2002.09293}{arXiv:2002.09293}]

\bibitem[\protect\citeauthoryear{Nunes \& Bernui}{2020b}]{Nunes2}
Nunes, R. C., Bernui, A., 2020b, 
\href{https://link.springer.com/article/10.1140/epjc/s10052-020-08601-8}
{Eur. Phys. J. C}, 80, 1025, 
[\href{https://arxiv.org/abs/2008.03259}{arXiv:2008.03259}]

\bibitem[\protect\citeauthoryear{Nunes \& Vagnozzi}{2021}]{Nunes21}
Nunes, R. C., Vagnozzi, S., 2021, \href{https://academic.oup.com/mnras/article-abstract/505/4/5427/6293862?redirectedFrom=fulltext}{MNRAS}, 505, 5427, [\href{https://arxiv.org/abs/2106.01208}{arXiv::2106.01208}]
 


\bibitem[\protect\citeauthoryear{Pandey}{2021a}]{Pandey21a}
Pandey, B., 2021, \href{https://iopscience.iop.org/article/10.1088/1475-7516/2021/02/023}{J. Cosmol. Astropart. Phys.}, 02, 023, [\href{https://arxiv.org/abs/2008.10266}{arXiv:2008.10266}]

\bibitem[\protect\citeauthoryear{Pandey \& Sarkar}{2021b}]{Pandey21b}
Pandey, B., Sarkar, S., 2021, \href{https://iopscience.iop.org/article/10.1088/1475-7516/2021/07/019}{J. Cosmol. Astropart. Phys.}, 07, 019,
[\href{https://arxiv.org/abs/arXiv:2103.11954}{arXiv:2103.11954}]

\bibitem[\protect\citeauthoryear{Pedregosa et al.}{2011}]{Pedregosa}
Pedregosa, F., et al., 2011, \href{https://www.jmlr.org/papers/volume12/pedregosa11a/pedregosa11a.pdf?source=post_page---------------------------}{J. Mach. Learn. Res.}, 12, 2825, [\href{https://arxiv.org/abs/1201.0490}{arXiv:1201.0490}]

\bibitem[\protect\citeauthoryear{Peebles}{1965}]{Peebles65}
Peebles, P. J. E., 1965, \href{http://articles.adsabs.harvard.edu/pdf/1965ApJ...142.1317P}{ApJ}, 142, 1317

\bibitem[\protect\citeauthoryear{Peebles}{1980}]{peebles80}
Peebles, P. J. E., 1980, The large-scale structure of the universe. Princeton Univ. Press, 
Princeton, NJ

\bibitem[\protect\citeauthoryear{Perivolaropoulos \& Skara, F.}{2021}]{Perivolaropoulos21}
Perivolaropoulos, L., Skara, F., 2021, [\href{https://arxiv.org/abs/2105.05208}{arXiv:2105.05208}]

\bibitem[\protect\citeauthoryear{Pezzotta et al.}{2017}]{Pezotta}
Pezzotta, A., et al., 2017, \href{https://www.aanda.org/articles/aa/abs/2017/08/aa30295-16/aa30295-16.html}{A\&A}, 604, A33, [\href{https://arxiv.org/abs/1612.05645}{arXiv:1612.05645}]

\bibitem[\protect\citeauthoryear{Pereira \& Pitrou}{2015}]{Pereira15}
Pereira, T., Pitrou, C., 2015, 
\href{https://www.sciencedirect.com/science/article/pii/S1631070515001528?via3Dihub}
{Comptes rendus - Physique}, 16, 1027, 
[\href{https://arxiv.org/abs/1509.09166}{arXiv:1509.09166}]

\bibitem[\protect\citeauthoryear{Planck Collaboration}{2020}]{Planck18}
Planck Collaboration, 2020, 
\href{https://www.aanda.org/articles/aa/abs/2020/09/aa33910-18/aa33910-18.html}{A\&A}, 641, 
A6, [\href{https://arxiv.org/abs/1807.06209}{arXiv:1807.06209}]




\bibitem[\protect\citeauthoryear{Rasmussen}{2003}]{Rasmussen}
Rasmussen, C. E., 2003, Gaussian processes in machine learning. In Summer school on machine learning, 63. Springer, Berlin, Heidelberg.

\bibitem[\protect\citeauthoryear{Renzi, \& Silvestri}{2020}]{Renzi20}
Renzi, F., Silvestri, A., 2020, [\href{https://arxiv.org/abs/2011.10559}{arXiv:2011.10559}]

\bibitem[\protect\citeauthoryear{\u{R}\'{\i}pa \& Shafieloo}{2019}]{Shafieloo}
\u{R}\'{\i}pa, J., Shafieloo, A., 2019 \href{https://academic.oup.com/mnras/article/486/3/3027/5480162}{MNRAS}, 486, 3027, [\href{https://arxiv.org/abs/1809.03973}{arXiv:1809.03973}]


\bibitem[\protect\citeauthoryear{Ross et al.}{2007}]{Ross07}
Ross, N. P., et al., 2007, \href{https://academic.oup.com/mnras/article/381/2/573/1018324}{MNRAS}, 381, 573, [\href{https://arxiv.org/abs/astro-ph/0612400}{arXiv:astro-ph/0612400}]


\bibitem[\protect\citeauthoryear{Sagredo, Nesseris, \& Sapone}{2018}]{Sagredo18}
Sagredo, B., Nesseris, S., Sapone, D., 2018, \href{https://journals.aps.org/prd/abstract/10.1103/PhysRevD.98.083543}{Phys. Rev. D}, 98, 083543, [\href{https://arxiv.org/abs/1806.10822}{arXiv:1806.10822}]

\bibitem[\protect\citeauthoryear{Schneider}{2014}]{Schneider}
Schneider, P., 2006, Extragalactic astronomy and cosmology: an introduction. Springer, Berlin

\bibitem[\protect\citeauthoryear{Scrimgeour et al.}{2012}]{Scrimgeour}
Scrimgeour, M. I., 2012, \href{https://academic.oup.com/mnras/article/425/1/116/996621}{MNRAS}, 425, 116, [\href{https://arxiv.org/abs/1205.6812}{arXiv:1205.6812}]

\bibitem[\protect\citeauthoryear{Seikel, Clarkson, \& Smith}{2012}]{Seikel12}
Seikel, M., Clarkson, C., Smith, M., 2012, \href{https://iopscience.iop.org/article/10.1088/1475-7516/2012/06/036}{J. Cosmol. Astropart. Phys.}, 06, 036, [\href{https://arxiv.org/abs/1204.2832}{arXiv:1204.2832}]


\bibitem[\protect\citeauthoryear{Shafieloo, Kim, \& Linder}{2012}]{Shafieloo12}
Shafieloo, A., Kim, A. G., Linder, E. V., 2012, \href{https://journals.aps.org/prd/abstract/10.1103/PhysRevD.85.123530}{phys. Rev. D}, 85, 123530, [\href{https://arxiv.org/abs/1204.2272}{arXiv:1204.2272}]

\bibitem[\protect\citeauthoryear{Silk}{1968}]{Silk68}
Silk, J., 1968, \href{http://articles.adsabs.harvard.edu/pdf/1968ApJ...151..459S}{ApJ}, 151, 
459


\bibitem[\protect\citeauthoryear{Song \& Percival}{2009}]{Song09}
Song, Y. S., Percival, W. J., 2009, \href{https://iopscience.iop.org/article/10.1088/1475-7516/2009/10/004}{J. Cosmol. Astropart. Phys.}, 10, 004, [\href{https://arxiv.org/abs/0807.0810}{arXiv:0807.0810}]

\bibitem[\protect\citeauthoryear{Strauss \& Willick}{1995}]{Strauss}
Strauss, M. A., Willick, J. A., 1995, \href{https://www.sciencedirect.com/science/article/abs/pii/0370157395000137?via\%3Dihub}{Phys. Rep.}, 261, 271, [\href{https://arxiv.org/abs/astro-ph/9502079}{astro-ph/9502079}]

\bibitem[\protect\citeauthoryear{Sunyaev \& Zeldovich}{1970}]{Sunyaev}
Sunyaev, R. A., Zeldovich, Y. B., 1970, \href{http://articles.adsabs.harvard.edu/pdf/1970Ap\%26SS...7....3S}{Ap\&SS}, 7, 3

\bibitem[\protect\citeauthoryear{Sun, Jiao, \& Zhang}{2021}]{Sun21}
Sun, W., Jiao, K., Zhang, T. J., 2021, [\href{https://arxiv.org/abs/2105.12618}{arXiv:2105.12618}]

\bibitem[\protect\citeauthoryear{Tarnopolski}{2017}]{Tarnopolski17}
Tarnopolski, M., 2017, 
\href{https://academic.oup.com/mnras/article/472/4/4819/4157286}{MNRAS}, 
472, 4819, 
[\href{https://arxiv.org/abs/1512.02865}{arXiv:1512.02865}]

\bibitem[\protect\citeauthoryear{Tegmark et al.}{2006}]{Tegmark06}
Tegmark, M., et al., 2006, \href{https://journals.aps.org/prd/abstract/10.1103/PhysRevD.74.123507}{Phys. Rev. D}, 74, 123507, [\href{https://arxiv.org/abs/astro-ph/0608632}{arXiv:astro-ph/0608632}]



\bibitem[\protect\citeauthoryear{Velasquez-Toribio, \& Fabris}{2020}]{Velasquez20}
Velasquez-Toribio, A. M.,  Fabris, J. C., 2020, \href{https://link.springer.com/article/10.1140\%2Fepjc\%2Fs10052-020-08785-z}{Eur. Phys. J. C.}, 80, 1, [\href{https://arxiv.org/abs/2008.12741}{arXiv:2008.12741}]







\bibitem[\protect\citeauthoryear{Zhang \& Li}{2018}]{Zhang}
Zhang, M. J., Li, H., 2018, \href{https://link.springer.com/article/10.1140/epjc/s10052-018-5953-3}{Eur. Phys. J. C}, 78, 1, [\href{https://arxiv.org/abs/1806.02981}{arXiv:1806.02981}]

\end{thebibliography}



\appendix

\section{Baryon Acoustic Oscillations' influence on the DPL approximation}\label{appendixA}

In section \ref{DPL} we have seen a discrepancy of 0.3\% between the fiducial model and the DPL model for $R_{\text{H}} \simeq 80$ Mpc/h. 
This small deviation in the fit can be attributed to the Baryon Acoustic Oscillations (BAO) 
signature, present around the scale $100$ Mpc/h. 
To test this hypothesis, we perform our fit for the DPL approximation considering two estimates of the correlation function:
one from the CLASS code and another for the case of absence of the BAO feature. 
For the last case, we use the fitting model given by~\cite{EH98} and implemented in the code 
{\tt nbodykit}~\footnote{\url{https://nbodykit.readthedocs.io/en/latest/}} to obtain $\xi(r)$.

\begin{figure}
\includegraphics[scale=0.5]{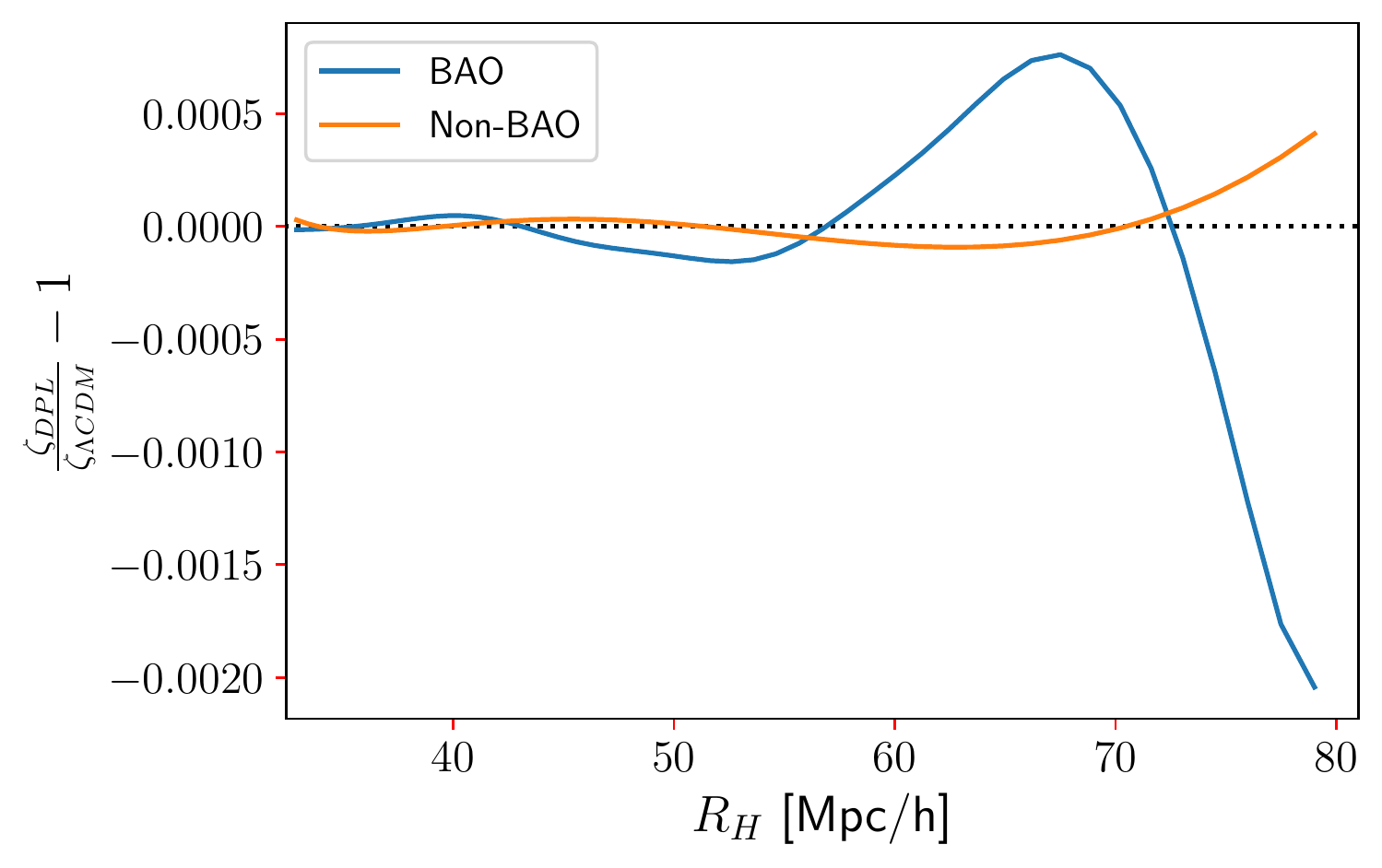}
\caption{
The relative difference $\zeta_{\text{DPL}}/ \zeta_{\Lambda\text{CDM}} -1$ obtained calculating 
the correlation function with and without the BAO feature.
}
\label{fig:withBAO}
\end{figure}

Figure \ref{fig:withBAO} shows the relative difference $\zeta_{\text{DPL}}/ \zeta_{\Lambda\text{CDM}} -1$ obtained calculating the correlation function with and without the BAO feature. 
It is evident the improvement in the fitting of $\zeta$ obtained from the correlation function without the BAO feature for 
$R_{\text{H}} < 40$ Mpc/h, with more significant effect for 
$R_{\text{H}} \gtrsim 55$ Mpc/h when compared to the correlation function with BAO.

\section{Studying the parameter dependence of the growth rate}\label{appendixB}

In section \ref{3.1} we found a relation between $f(z)$ and $R_{\text{H}}(z)$ that is, in principle, 
explicitly independent of the $\epsilon$ parameter. However, when 
measuring  $R_{\text{H}}(z)$, one needs to fix $\epsilon$. Then, it is important to check if this criterion affects the $f(z)$ estimate. 
Here we investigate the impact of fixing $\epsilon$, as well as whether our choice of cosmological parameters might affect the $f(z)$ obtained.

We compare the result obtained from our fiducial model,  $f^{\text{fiducial}}(z)$, using the input parameters 
\begin{equation}\label{paramfiducial}
\{\epsilon, \ln(10^{10}A_{s}),\Omega_{K}, \Omega_{c}h^{2}\} = \{0.01, 3.045, 0.0, 0.1202\},
\end{equation}
with the $f(z)$ resulting  from the same fitting procedure but now varying these four parameters one at a time. The comparison is performed 
through the relative difference $f(z)/f^{\text{fiducial}}(z) - 1$.

\begin{figure}
\includegraphics[scale=0.5]{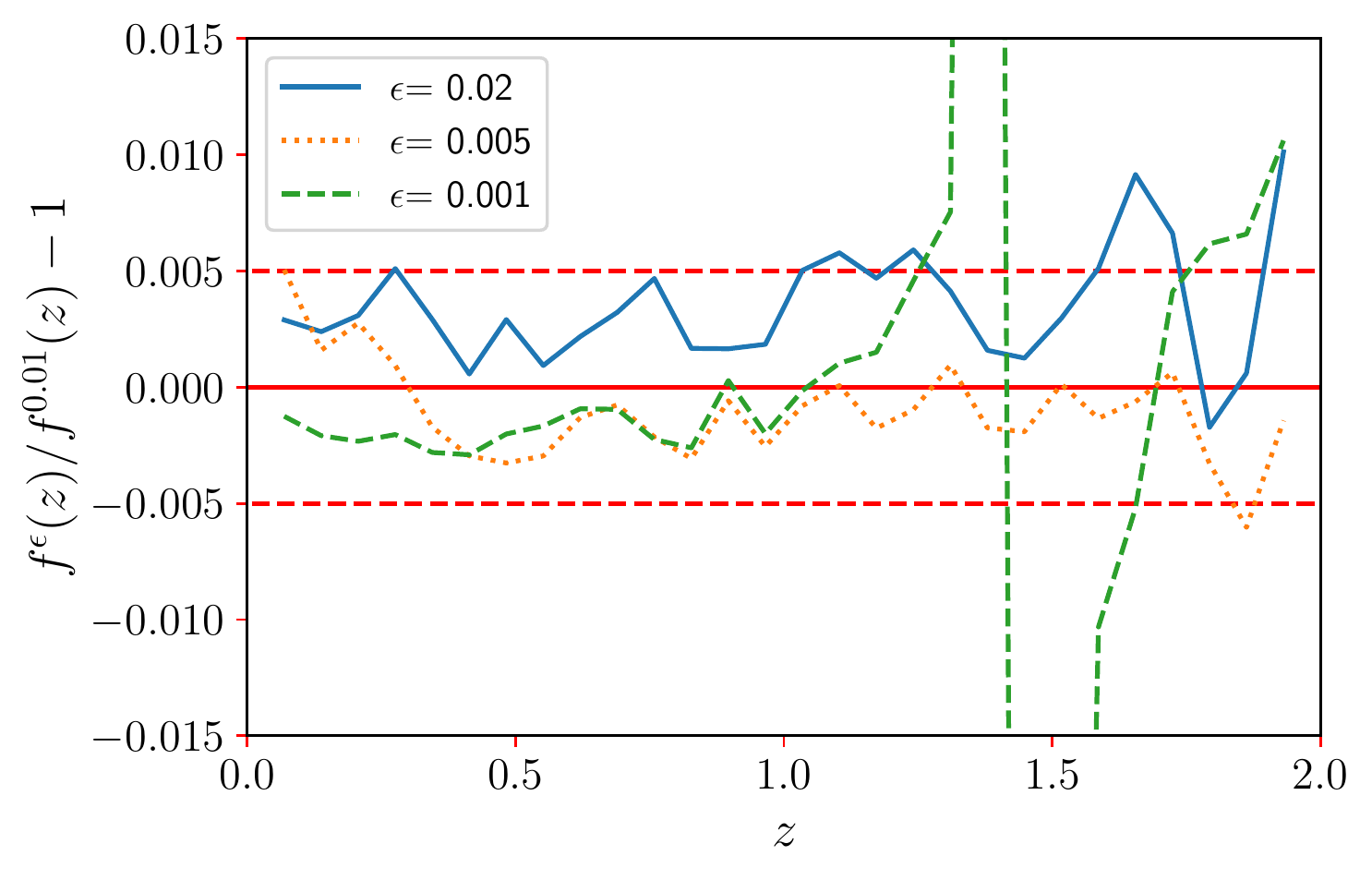}
\caption{ 
The relative difference for $f(z)$ between our fiducial model, i.e., $ \epsilon=0.01$, and the 
cases investigated with $\epsilon=\{0.02,0.005,0.001\}$.}
\label{fig:frecases}
\end{figure}

Figure \ref{fig:frecases} displays the relative difference between our input model 
and the 3 cases studied where $\epsilon = \{0.02, 0.05, 0.001\}$. 
These values correspond to different definitions of the homogeneity scale, $R_{H}$, where this 
scale is obtained when the data in analysis reaches 2\%, 0.5\%, and 0.1\% below the limit 
value 3, respectively~\citep{Scrimgeour}. 
We show that for the redshift interval of interest, $0 < z < 2$, the error is below $0.5\%$, 
which makes our approach robust with respect to $\epsilon$. 
For the case $\epsilon=0.001$, 
a divergent behaviour is observed around $z \sim 1.4$, where the function explodes up and 
come back from below. 
However, we notice that such small $\epsilon$ is unpractical when investigating 
$R_{H}$ due to the statistical errors (and other systematics) inherent to the data analyses.

\begin{figure*}
\mbox{
\includegraphics[scale=0.4]{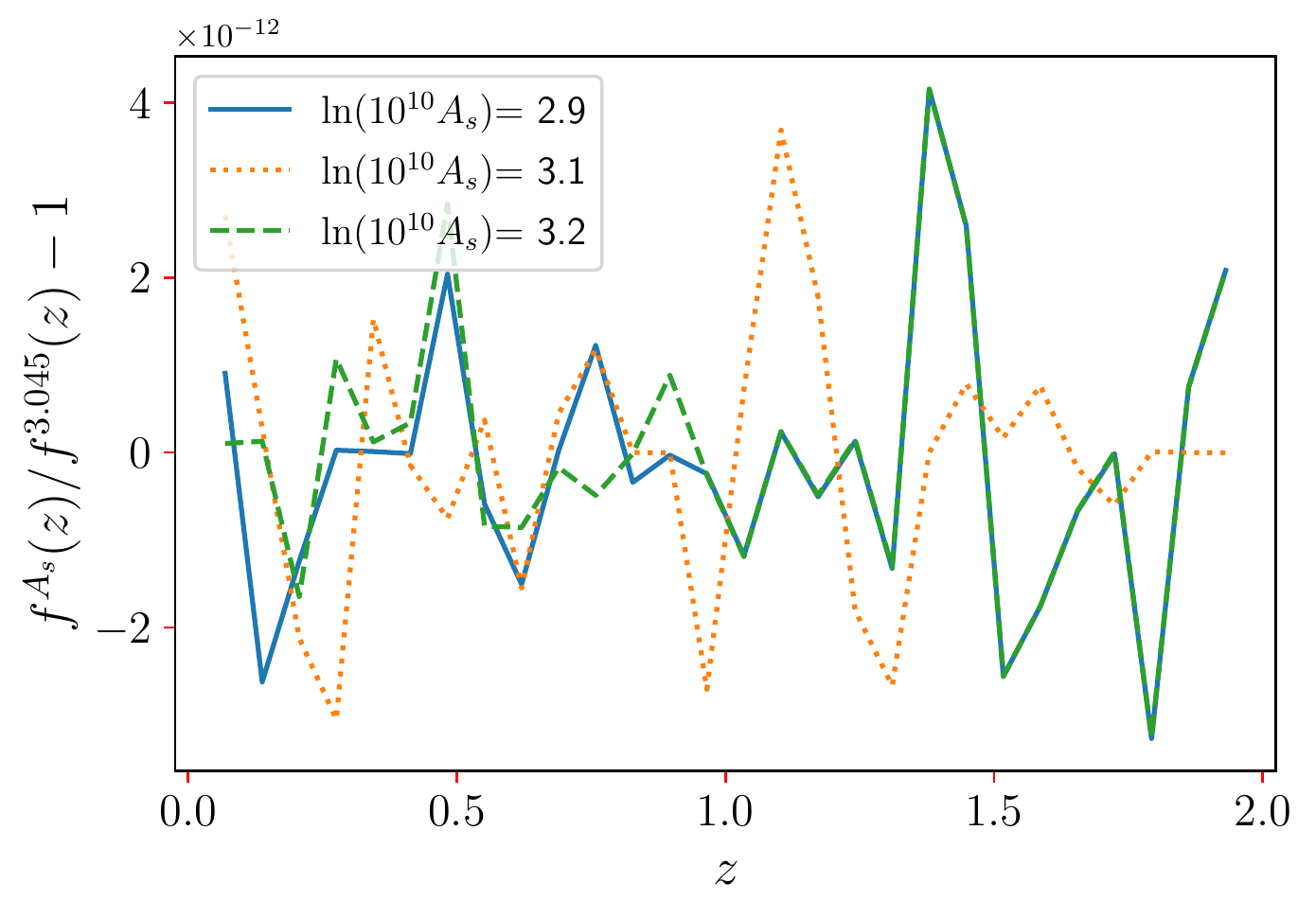} 
\includegraphics[scale=0.4]{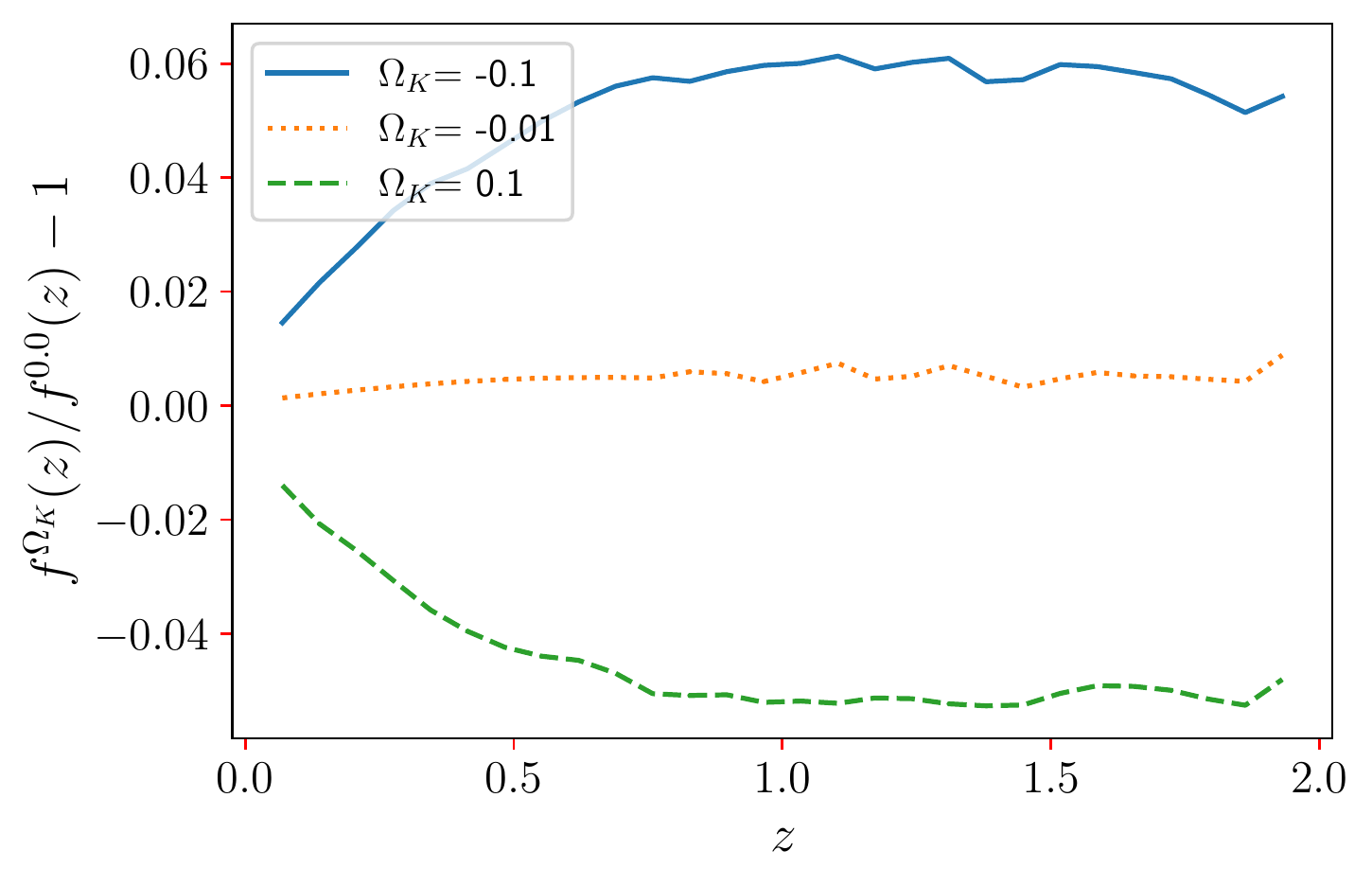} 
\includegraphics[scale=0.4]{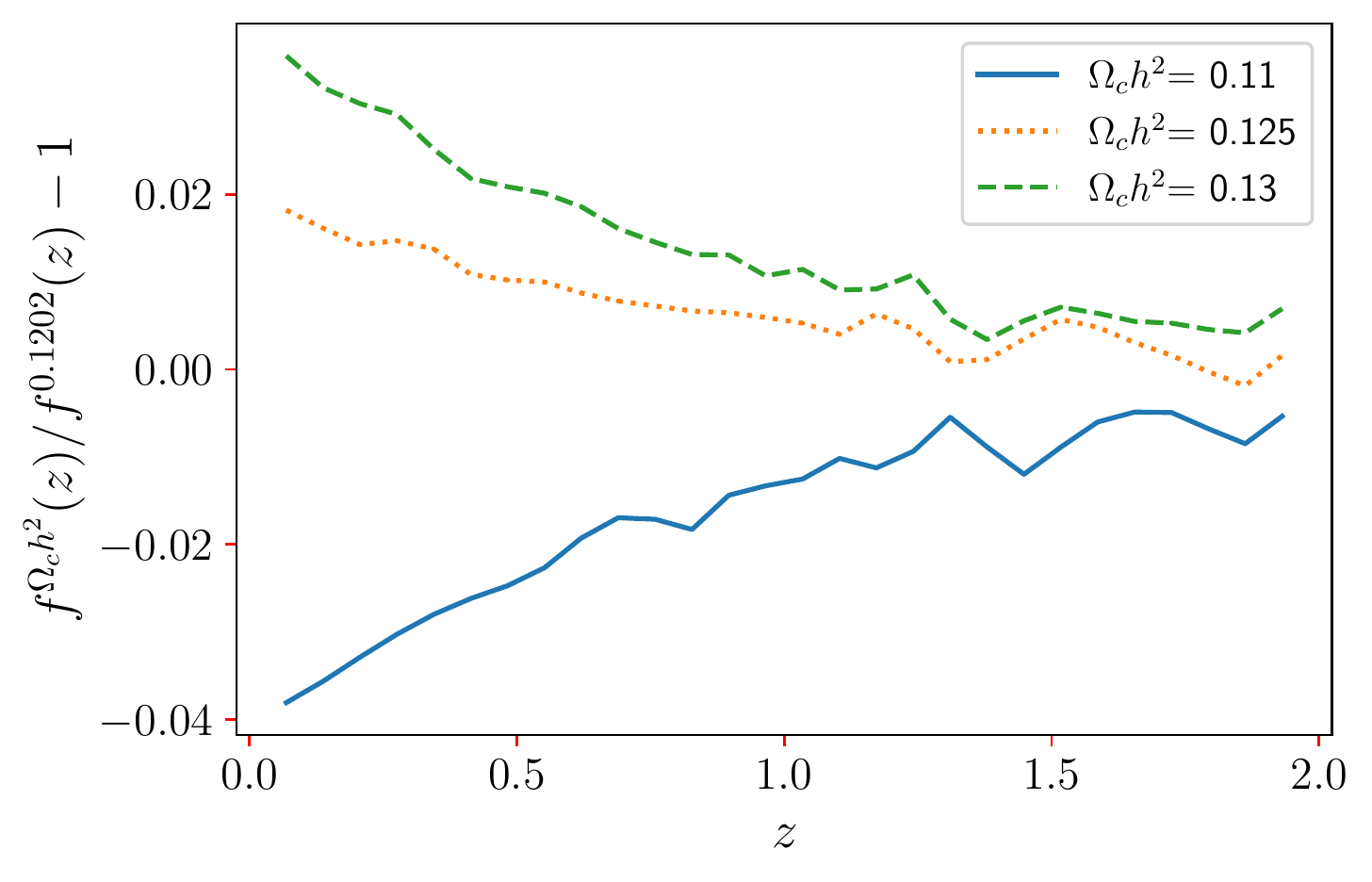}
}
\caption{The relative difference, $f(z)/f^{\text{fiducial}}(z) - 1$, 
considering the variations, one at a time, of three cosmological parameters: 
$\ln(10^{10}A_{s}), \Omega_{K}$, and $\Omega_{c}h^{2}$. 
The left panel shows the results for $A_{s}$, which is basically noise. 
The middle panel shows the outcomes for $\Omega_{K}$ where the relative difference is less 
than $6\%$ for the whole interval of interest. 
The right panel shows the dependence on  $\Omega_{c}h^{2}$ which is $\lesssim 4\%$, with the 
largest values for low $z$.}
\label{fig:fracases}
\end{figure*}

Figure \ref{fig:fracases} shows $f(z)/f^{\text{fiducial}}(z) - 1$ for analyses 
obtained with the variation of three  cosmological parameters one at a time: 
$\ln(10^{10}A_{s}), \Omega_{K}$, and $\Omega_{c} h^{2}$. 
For $\ln(10^{10}A_{s})$ we use $\{2.9, 3.1, 3.2\}$, which is a large enough interval when we 
compare with the Planck best-fit, namely, $\ln(10^{10}A_{s}) = 3.045 \pm 0.016$. 
Our results, displayed in the left panel of figure~\ref{fig:fracases}, show nothing but statistical 
noise, indicating that our model is independent of $\ln(10^{10}A_{s})$ values. 
For $\Omega_{K}$ we consider $\{-0.1, -0.01, 0.1\}$ (see figure~\ref{fig:fracases}, middle panel). 
Also well beyond $2\sigma$ uncertainty for the Planck best-fit $\Omega_{K}= -0.044 \pm 0.0165$. 
For all these cases, we observe a maximum of 6\% deviation, for the whole redshift interval. 
For all purposes, our approach has a small dependence on $\Omega_{K}$ considering 
a large interval of possible values. 

For the analyses of the last parameter, $ \Omega_{c} h^{2} $, we consider 
$\{0.11, 0.125, 0.13\}$. 
In these cases, we also find a slight dependence in our results, $\lesssim 4\%$, and 
decreasing for high $z$ (see the right panel of figure~\ref{fig:fracases}). 
As in the previous analyses, this result was somehow expected, because we are not modifying 
the meaning of $f(z)$, we just found an alternative way to find it. 
We already knew that the growth rate has a strong dependence in the matter density parameter, 
as seen in the parametrization $f(z) = \Omega_m(z)^{\gamma}$, where $\gamma$ depends 
only on the constant of the 
equation of state, $\omega=-1$, for the $\Lambda$CDM case, or 
modifications according to the gravity model used.

\section{Consistency test for different kernels}\label{appendixC}

Our main result, i.e., the reconstruction of the homogeneity scale function, which, in turn, we use to derive the growth rate of structures, 
is based on the SE kernel. 
The SE kernel is a smooth covariance function that can reproduce global characteristics, although 
sometimes it cannot reproduce local characteristics. 

As our data sample is nicely distributed, this kernel works smoothly. 
In order to test for possible systematic effects on the kernel choice, we examine our main results using this time the Mat\'ern class kernels. 
The Mat\'ern kernel can be written as 
\begin{equation}
K_{M_{\nu}}(\tau) = \sigma_f^2 \,\frac{2^{1-\nu}}{\Gamma(\nu)}\, \left( \frac{\sqrt{2 \nu}\,\tau}{l} \right)^{\nu} K_{\nu}\left( \frac{\sqrt{2 \nu}\,\tau}{l} \right),
\end{equation}
where $K_{\nu}$ is the modified Bessel function of second kind, $\Gamma(\nu)$ is the standard Gamma function and $\nu$ is 
a strictly positive parameter. Here, the hyper-parameters $\sigma_f$ and $l$ are also optimized during the fitting.

\begin{figure}
\centering
\includegraphics[scale=0.6]{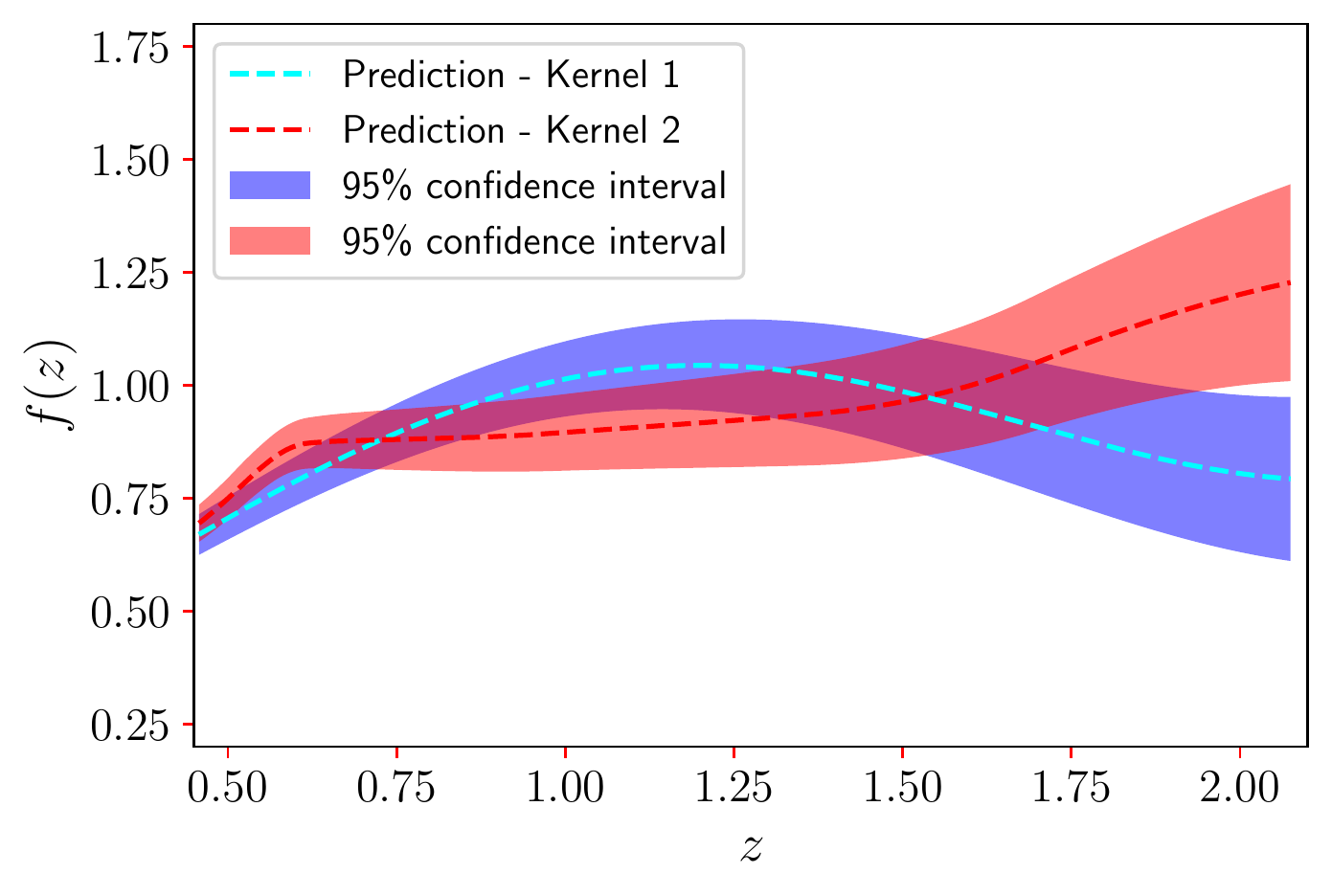}
\caption{The reconstruction of the growth rate of structures $f^{R_{\text{H}}}(z)$ 
using two different kernels, namely, kernel 1 (the SE kernel) and kernel 2 (the 
Mat\'ern kernel).}
\label{fig:k2}
\end{figure}

Figure \ref{fig:k2} shows the best-fit prediction and the GP reconstruction of the $f^{R_{\text{H}}}(z)$ function using the SE and the Mat\'ern kernels. 
We do not find significant differences between both reconstructions, therefore, we conclude that they are statistically equivalent.

\section{Measurements of the growth rate function}\label{appendixD}

The literature reports diverse compilations of measurements of the growth rate of cosmic 
structures, $f(z)$ \cite[see, e.g.][]{Basilakos12, Nunes16,Sagredo18}, which we update here. 
Our compilation, shown in  Table~\ref{tableD1}, follows three criteria in order to avoid or minimize possible data correlations, 
we consider: (i) $f(z)$ measurements obtained  with cosmic tracers from  different astronomical surveys or from disjoint redshift bins;       
(ii) direct measurements of $f$, and not measurements of 
$f \sigma_8$ that use a fiducial cosmological model to eliminate the $\sigma_8$-dependence;
(iii) the latest measurement of $f$ when the same astronomical survey performed two or more measurements corresponding 
to several data releases.

\begin{table*}
\centering
\setlength{\extrarowheight}{0.11cm}
\caption{Data compilation of $f(z)$ measurements that shares important features, as explained in the Appendix~\ref{appendixD}.
}
\label{tableD1}
\begin{tabular}{|c|c|c|c|c|}
\hline
Survey &$z$ & $f$ & Reference & Cosmological tracer \\
\hline
ALFALFA & 0.013 & $0.56 \pm 0.07$ & \cite{Avila21} & HI extragalactic sources \\
2dFGRS & 0.15 & $0.49 \pm 0.14$ & \cite{Hawkins03,Guzzo08} & galaxies \\
GAMA & 0.18 & $0.49 \pm 0.12$ & \cite{Blake13} &  
multiple-tracer: blue \& red gals. \\
WiggleZ   & 0.22 & $0.60 \pm 0.10$ & \cite{Blake11} & galaxies \\
SDSS    & 0.35 & $0.70 \pm 0.18$ & \cite{Tegmark06} & luminous red galaxies (LRG) \\
GAMA   & 0.38 & $0.66 \pm 0.09$ & \cite{Blake13} & 
multiple-tracer: blue \& red gals. \\
WiggleZ & 0.41 & $0.70 \pm 0.07$ & \cite{Blake11} & galaxies \\
2SLAQ & 0.55 & $0.75 \pm 0.18$ & \cite{Ross07} & LRG \& QSO\\
WiggleZ & 0.60 & $0.73 \pm 0.07$ & \cite{Blake11} & galaxies \\
VIMOS-VLT Deep Survey & 0.77 & $0.91 \pm 0.36$ & \cite{Guzzo08} & faint galaxies \\
2QZ \& 2SLAQ & 1.40 & $0.90 \pm 0.24$ & \cite{DaAngela08} & QSO \\
\hline
\end{tabular}
\end{table*}

%
%


\bsp	
\label{lastpage}
\end{document}